\documentclass[pra,aps,twocolumn,preprintnumbers,showpacs,tightenlines,nofootinbib,10pt]{revtex4-1}
\usepackage{latexsym,epsfig,amssymb,amsfonts,amsmath,graphicx,bbm,mathptm,hyperref}
\pdfoutput=1 
\usepackage{mathtools}   					
\usepackage{bbold}
\usepackage{csquotes}
\usepackage{color}						
	\definecolor{darkblue}{rgb}{0,0,.5}
	\definecolor{black}{rgb}{0,0,0}

\allowdisplaybreaks

\begin{document}

\title{On Baryogenesis and $\mathbf{n\bar n}$-Oscillations}

\author{Enrico \surname{Herrmann}}
\email[email: ]{eherrmann@caltech.edu}
\affiliation{ Walter Burke Institute for Theoretical Physics,\\
California Institute of Technology, Pasadena, CA 91125 }%
\date{\today}
\preprint{CALT-TH-2014-153}

\begin{abstract}
\noindent
We study a simple model where color sextet scalars violate baryon number at tree level but do not give rise to proton decay. In particular, we include one light and two heavy sextets with $\Delta B=2$ baryon number violating interactions that induce neutron anti-neutron oscillations. This setup also suggests an intimate connection to the generation of the observed baryon asymmetry in the Universe via the out of equilibrium decay of the heavy sextet scalars at around $10^{14}$ GeV. The large $SU(3)$-color charges of the scalar fields involved in generating the baryon asymmetry motivate us to study potentially significant washout effects. We numerically solve a set of Boltzmann evolution equations and find restrictions on the available model parameters imposed by successful high scale baryogenesis. Combining our new numerical results for baryogenesis with $n\bar n$-oscillation predictions and collider limits on the light sextet, we identify parameter regions where this model can be probed by current and future experiments. 
\end{abstract}


\maketitle

\section{Introduction}

The origin of the baryon asymmetry in the Universe (BAU) is one of the major open puzzles not solved within the Standard Model of particle physics.
Over the years, a wide range of models and mechanisms to generate the observed baryon asymmetry have been suggested. 
Electroweak baryogenesis \cite{Cohen:1993nk}, baryogenesis via leptogenesis \cite{Luty:1992un}, the Affleck-Dine mechanism \cite{Affleck:1984fy} and GUT-baryogenesis \cite{Kolb:1979qa} 
are some of the most common representatives. For a comprehensive review, we encourage the reader to consult e.g.~\cite{Riotto:1998bt} and references therein.
\newline\newline
In 1967, Sakharov \cite{Sakharov:1967dj} formulated three necessary ingredients for successful baryogenesis: 1) $CP$-violation, 2) baryon number violation and 3) the deviation from thermal equilibrium. 
Perturbatively, the Standard Model preserves baryon-number, but this accidental global symmetry is anomalous at the quantum level. In principle, the Standard Model contains all required ingredients to generate a baryon asymmetry. However it was shown that the attainable asymmetry is inconsistent with observation \cite{Huet:1994jb}. 
Beyond the Standard Model there are a variety of well motivated extensions (e.g.\cite{Georgi:1974sy,Dimopoulos:1981zb,Pati:1974yy}) that introduce new interactions which violate baryon number classically. Generically, this leads to proton decay which is tightly constrained by experiment \cite{Nishino:2009aa}. 
In this article, we discuss a minimal model, suggested in \cite{Arnold:2012sd}, that violates baryon number in such a way that the proton decay channel is absent and $n\bar n$-oscillations become the primary signal of baryon number violating physics. In particular, we consider the situation where one light ($X_1$) and one heavy ($X_2$) color sextet scalar field mediate these interactions.  
\newline\newline
One attractive feature of this scenario is the possibility to integrate our simplified model into a complete non-supersymmetric grand unified theory in the spirit of Ref.~\cite{Babu:2012vc}. There, one of the color sextets is kept at the TeV-scale to ensure gauge coupling unification.
\newline\newline
Building on the discussions in Refs.~\cite{Arnold:2012sd,Babu:2012vc} we show that the model is able to accommodate successful high-scale baryogenesis 
as well as $n\bar n$-oscillations with discovery potential at the European Spallation Source (ESS)\footnote{\url{http://europeanspallationsource.se/fundamental-and-particle-physics}} \cite{Peggs:2009zza}. 
However the preferred parameter regions for the two effects are in some tension (c.f. results in Sec.~\ref{subsec:NumResBaryogenesis} and Sec.~\ref{sec:NNbarOscillations}). 
We perform a detailed analysis of the baryogenesis setup beyond the approximate decay scenarios 
assumed in the literature \cite{Arnold:2012sd,Babu:2012vc} by numerically solving the Boltzmann equations to track 
the relevant particle densities in the early Universe. This way, we correctly take into account potentially large washout contributions due to the 
sizable color charge of the new scalars. We also consider collider limits on new colored particles from LHC-searches that have been analyzed for color sextets in Ref.~\cite{Richardson:2011df}.
\newline\newline
The remainder of this note is structured as follows: 
In section \ref{sec:Model} we introduce the field content and the parameters of our model which are relevant for the discussion of baryogenesis and $n\bar n$-oscillations. 
In section \ref{sec:Baryogenesis} we rehash some facts about baryogenesis before writing down the Boltzmann equations in subsection \ref{subsec:BoltzmannEq}. 
We also compute the relevant scattering- and decay-rates that feed into these equations.   
Subsequently, we discuss $n\bar n$-oscillations in section \ref{sec:NNbarOscillations} and show how to combine this with successful baryogenesis. This allows us to analyze the the two phenomena in terms of physically motivated quantities and measurable parameters only.
Additional signals and constraints of our model, such as collider signatures, contributions to the neutron electric dipole moment (Sec.~\ref{sec:EDMLHC}) as well as meson anti-meson mixing are briefly addressed. 
A detailed account of our conventions is deferred to appendix \ref{app:Boltzmann}.
%
%
\section{The Model}
\label{sec:Model}
%
%
We study one particular model with baryon-number violation but no proton decay, originally suggested in \cite{Arnold:2012sd}. In a similar context, Babu et al.~\cite{Babu:2012vc} discussed the role of color sextet scalars for non-supersymmetric gauge-coupling unification in a complete $SO(10)$-theory, $n\bar n$-oscillations and baryogenesis. 
Extending the analysis in \cite{Arnold:2012sd,Babu:2012vc}, we sharpen the link between $n\bar n$-oscillations and baryogenesis. For our purposes, we focus on a simplified model with three additional scalars. The relevant degrees of freedom and their respective charges under $G_{SM}=SU(3)_c \times SU(2)_L \times U(1)_Y$ summarized in table \ref{tab:fieldcontent}.
	 \begin{table}[ht!]
	 \centering
	    \begin{tabular}{|c || c|c|c|}
		\hline
		Field	&  $SU(3)_c$		& $SU(2)_L$ 	& $U(1)_Y$ 	\\
	       \hline
	       $X_1$	&	$\bar6$		& 	1	&  -1/3		\\
	       $X_2$	&	$\bar6$		& 	1	&  2/3		\\
	       $\widetilde X_2$	&	$\bar6$		& 	1	&  2/3		\\
	       $Q_L$	&	3		&	2	&  1/6		\\
	       $u_R$	&	3		&	1	&  2/3		\\
	       $d_R $	&	3		&	1	& -1/3		\\
	        \hline
	    \end{tabular}	    
	    \caption{\label{tab:fieldcontent}Field content and respective representation under the Standard Model gauge group. 
			$Q_L$ represents a left-handed quark doublet and $u_R,\ d_R$ are the right handed up- and down-type quarks. 
			}
	    \end{table}
The part of the color sextet Lagrangian relevant for baryogenesis and $n\bar n$-oscillations takes the form\footnote{We do not display the complete scalar potential nor parts of the Lagrangian involving the 
$\widetilde X_2$ field, required to generate $CP$-violation. In terms of an $SO(10)$-symmetric GUT-model there can be relations between different Yukawa couplings \cite{Babu:2012vc}.}:
\begin{align}
\label{eq:LagSextet}
 \mathcal{L}_{\text{sextet}} \supset & -g^{ab}_1 X^{\alpha\beta}_{1}(Q^a_{L\alpha}\epsilon Q^{b}_{L\beta}) - g^{ab}_2 X^{\alpha\beta}_2(d^a_{R\alpha}d^b_{R\beta}) \nonumber \\
	       & -g'^{ab}_1 X^{\alpha\beta}_1(u^a_{R\alpha}d^{b}_{R\beta})+ \lambda X^{\alpha\alpha'}_1 X^{\beta\beta'}_1 X^{\gamma\gamma'}_2 \epsilon_{\alpha\beta\gamma}\epsilon_{\alpha'\beta'\gamma'}
\end{align}
In eq.(\ref{eq:LagSextet}), letters $a, b,$ etc. denote the flavor structure, whereas greek letters $\alpha, \beta, \cdots$ represent $SU(3)$-color indices. 
Note that integrating out $X_2$, which we assume to be heavy, would generate a quartic coupling for $X_1$. 
This leads us to expect that the dimensionful coupling $\lambda$ is of order $M_2$, the mass of $X_2$.
\newline
Scalar sextets are represented as symmetric $3\times3$-matrices in color space,
\begin{align}
 \label{eq:ColorSextetRep}
 (X^{\alpha \beta}) = & \frac{1}{\sqrt{2}}
			\begin{pmatrix}
                         \sqrt{2} \tilde X^{11} 	& \tilde X^{12}			& \tilde X^{13} \\
                         \tilde X^{12}			& \sqrt{2} \tilde X^{22} 	& \tilde X^{23} \\
                         \tilde X^{13}			& \tilde X^{23}			& \sqrt{2} \tilde X^{33}
                        \end{pmatrix}.
\end{align}
Group theoretic details and color-flow Feynman rules for color sextets are given in the appendices of Refs.~\cite{Han:2009ya,Kilian:2012pz}. 
\newline
As written above, our model contains a vast number of free parameters. 
In the following, we restrict our discussion to the case, where the color sextet scalars only couple to the first generation				
of quarks. This reduces the number of Yukawa-couplings considerably. 
In order to generate a baryon asymmetry via $X_2$-decays (c.f. Feynman diagrams in fig.\ref{fig:X2X1X1Tree}), 
we have to introduce a second heavy sextet, $\widetilde X_2$ with a Lagrangian analogous to eq.\ref{eq:LagSextet} and couplings denoted by tildes. 
In Ref.~\cite{Arnold:2012sd} it was shown that even when $X_2$ and $\widetilde X_2$ couple to one generation of 
quarks only, it is impossible to remove all $CP$-violating phases of the model by field redefinitions.
The presence of $CP$-violation allows us to satisfy one of Sakharov's criteria so that $X_2$-decays prefer matter over anti-matter. We will comment on the exact form and the relevant size of the $CP$-violating parameter later.
\newline
For the color sextets, we neglect all interactions in the scalar potential besides the cubic $\lambda,\tilde\lambda$-terms. We choose $\lambda,\tilde\lambda$ to be real and move the phases to the Yukawa-couplings $g_2$ and $\tilde g_2$. 
In the discussion on $n\bar n$-oscillations, we also neglect $g_1$, the coupling of $X_1$ to the left-handed quarks, so that we are left with the following set of parameters
$\{M_1,M_2,\widetilde M_2, g_2,\tilde g_2, \lambda, \tilde \lambda,g'_1\}$.
%
%
\section{Baryogenesis}
\label{sec:Baryogenesis}
%
%
As discussed in the introduction, Sakharov \cite{Sakharov:1967dj} formulated 
three conditions for successful baryogenesis which we will discuss in the context of 
our model.  
\newline
The standard assumption \cite{Kolb:1990vq} of a baryon-symmetric big bang and the observational 
fact that there is more matter than anti-matter in the present Universe requires 
that baryon number has to be violated \cite{Sakharov:1967dj}. 
This is Sakharov's second condition. Nonperturbativly, the Standard Model has a source of 
baryon number violation in the form of instanton interactions \cite{'tHooft:1976fv,Arnold:1987mh}, 
which play a role in several baryogenesis scenarios, c.f. e.g. \cite{Cohen:1993nk,Riotto:1998bt}.
However, we focus on a model that explicitly breaks $B$ (and thereby also $(B-L)$) at tree level.
\newline
From the Lagrangian (\ref{eq:LagSextet}), it is easy to understand how this breaking takes place in our model. We envision a situation where one of the color sextets, $X_1$, is much lighter\footnote{An additional 
light fundamental scalar aggravates the fine-tuning problem of the Standard Model which we do not address here. In the GUT-framework suggested in Ref.\cite{Babu:2012vc} it is preferred to have a light sextet to help with gauge coupling unification.} then 
the other two, $X_2$ and $\widetilde X_2$. For the purpose of baryogenesis, we will treat $X_1$ as a stable particle ($g_1, g'_1 \ll1$). 
Its interactions with quarks dictate that $X_1$ has baryon number $-2/3$ and eventually decays hadronically.
Looking at the $g_2 X_2 d_{R}d_{R}$-term in eq.(\ref{eq:LagSextet}), we would also assign baryon number $-2/3$ to $X_2$, which makes it
impossible to consistently assign baryon number in the $\lambda X_1 X_1 X_2$-term
so that $B$ is violated by $X_2$-interactions.
\newline
We use a standard out of equilibrium decay of the heavy $X_2$-scalars to satisfy Sakharov's last condition. 
In an expanding Universe, processes go out of equilibrium (freeze out), when their rate is small compared to the expansion rate dictated by the Hubble parameter $H$. 
High scale baryogenesis scenarios became somewhat unfashionable when people realized the tension between high reheating temperatures required for these models and 
predictions within simple inflationary frameworks \cite{Riotto:1998bt}. 
However, if one takes into account a mechanism termed \textit{preheating} \cite{Greene:1997fu}, GUT-scale reheating temperatures can be achieved.
\subsection{Boltzmann equations}
\label{subsec:BoltzmannEq}
\begin{figure}[ht!]
    \centering
    \begin{align*}
    & \includegraphics[width=0.12\textwidth]{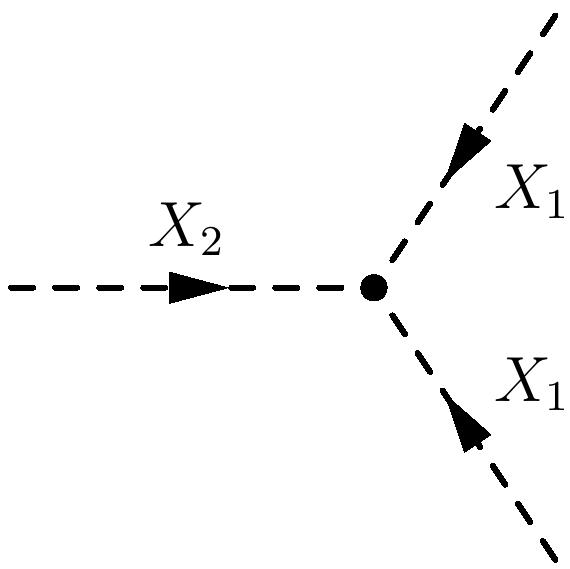} \quad \quad
    \includegraphics[width=0.20\textwidth]{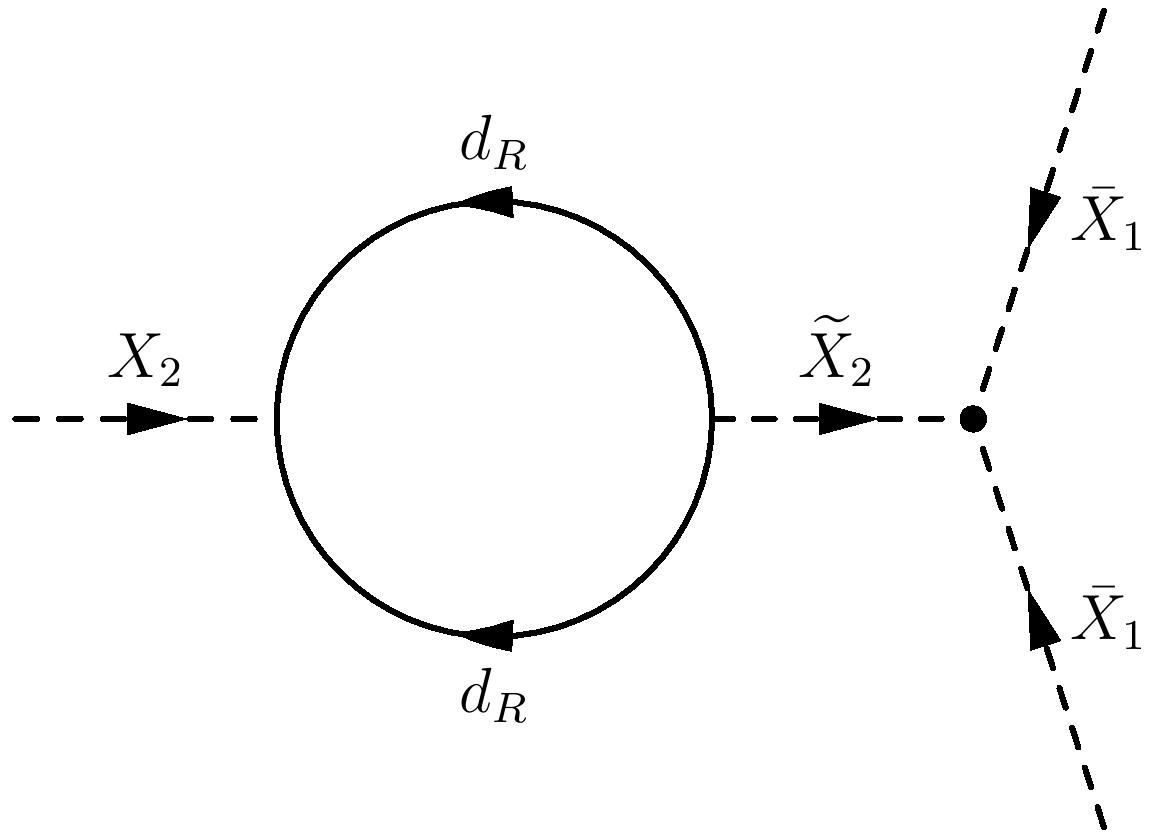} \\
    & \includegraphics[width=0.12\textwidth]{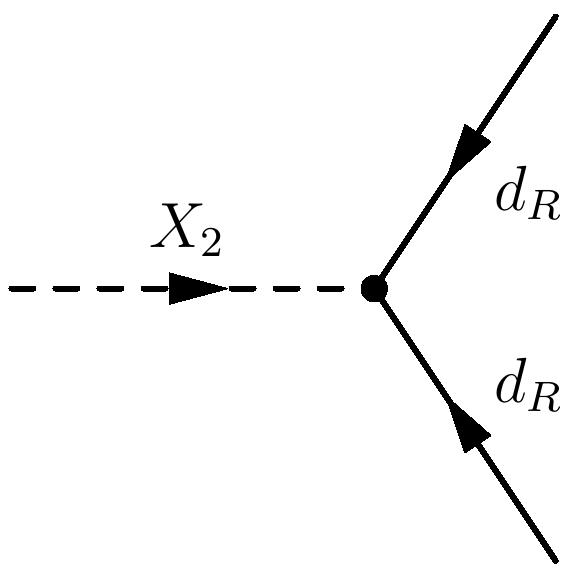} \hspace{1.5cm}
    \includegraphics[width=0.12\textwidth]{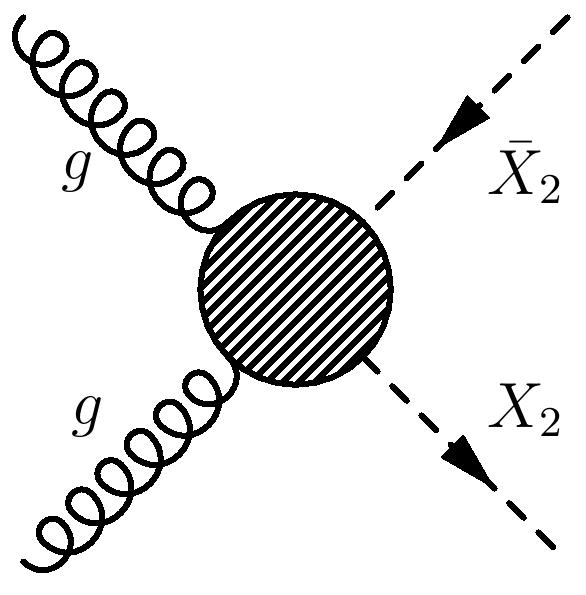}
    \end{align*}
    \caption{Top: Baryon number violating $\Delta B =2$ decays of $X_2$ at tree and one-loop level. \newline
             Bottom: We neglect a diagram with $X_1$ running in the loop, which is suppressed for $\lambda/M_2, \tilde\lambda/M_2 \ll g_2, \tilde g_2$. 
             Tree level $2\to2$-scattering processes between gluons and $X_2$s which could contribute to washout effects, by keeping $X_2$ in equilibrium, are taken into account.}
    \label{fig:X2X1X1Tree}
\end{figure}

\begin{figure}[!ht]
      \centering
	    \includegraphics[width=0.4\textwidth]{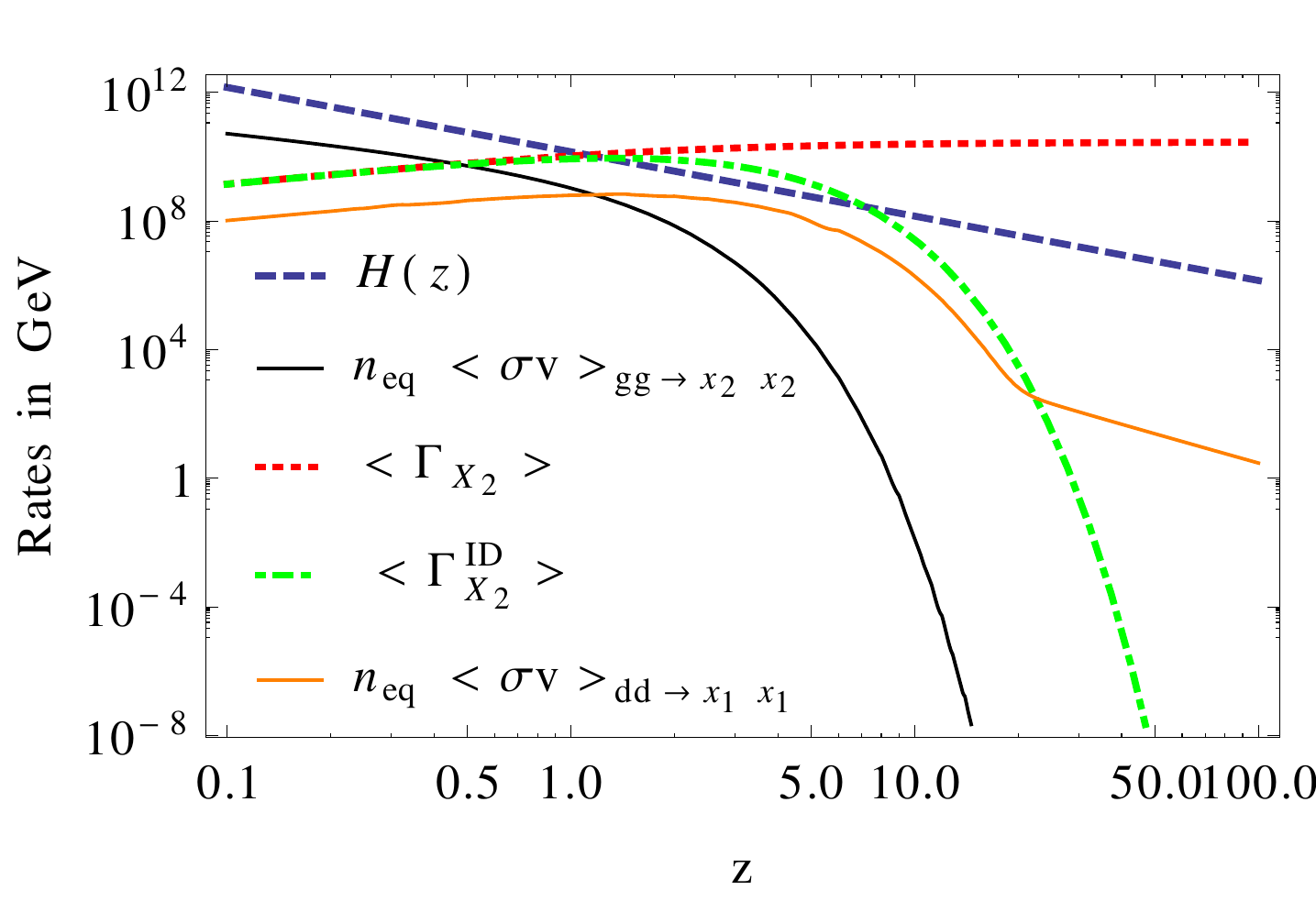}
        \caption{For one representative parameter point, we show the relevant decay ($\Gamma_{X_2}$), inverse decay ($\Gamma^{ID}_{X_2}$) and scattering rates that enter into the Boltzmann equations. We chose $M_2 =10^{14}$ GeV, 
		$\lambda = 3.5\times 10^{-2} M_2$ and $|g_2|^2 = 8.4 \times 10^{-2}$. Gluon scattering is not important and the dominant washout contribution comes from inverse decays that are active up to $z=$5-10.
		 Varying these parameters can shift the relevant rates with respect to the Hubble parameter $H(z)$. }
        \label{fig:BoltzmannRates}
\end{figure}

\begin{figure}[!ht]
      \centering
	    \includegraphics[width=0.32\textwidth]{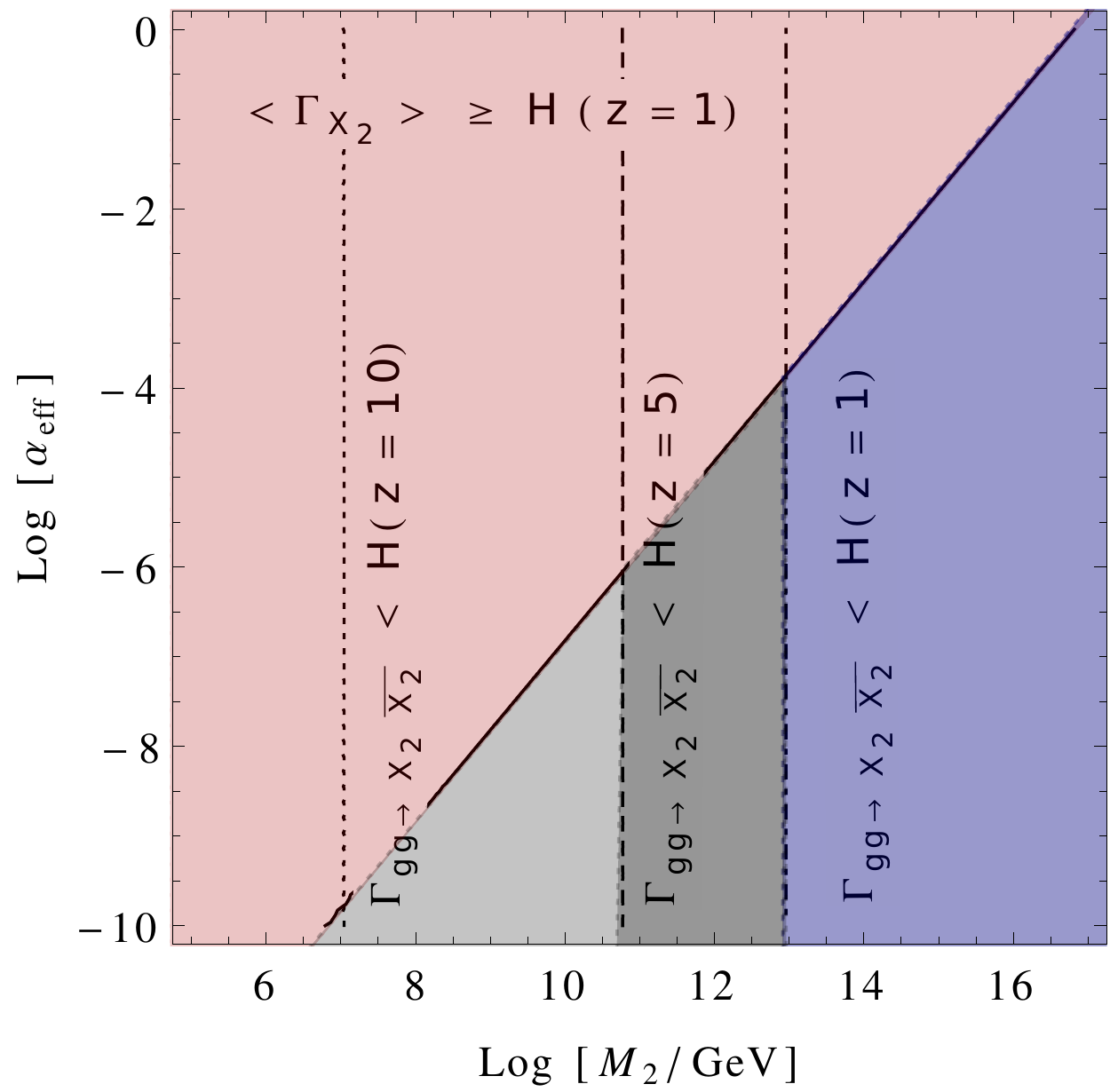}
        \caption{Estimate of allowed mass range and coupling strength for $X_2$ from baryogenesis considerations.
		  \newline
                  Contours indicate $\left.\langle\Gamma_{X_2}\rangle \right|_{z=1}=H(z=1)$ 
                  and $\Gamma_{gg\to X_2 \bar X_2} \equiv \left. n^{eq}_{g} \langle v\sigma(gg\to X_2 \bar X_2)\rangle \right|_{z=z^*}=H(z^*=1,5,10)$, where 
                  the thermally averaged reaction rates equal the Hubble rate. The red region represents parameter points where the $X_2$-decay rate is dominant.
                  The effective decay strength $\alpha_{\text{eff}}$ is related to our model parameters via eq.~\ref{eq:DefAlphaEff}.
                 }
        \label{fig:outofeqRatesContour}
\end{figure}

Boltzmann equations are a standard tool to accurately study the evolution of distribution functions and number densities in the early Universe. 
This is particularly important in parameter regions where 
simplified assumptions of a free evolution fail. 
In our model, the fields involved in the generation of the baryon asymmetry have a large color charge under the strong interaction which can lead to 
sizable reaction rates in comparison to the Hubble parameter. 
In order to explore the relation of successful baryogenesis and visible $n \bar n$-oscillations reliably, we solve a coupled system of Boltzmann equations for the abundance of $X_2$ and the light species $d,\ X_1$ numerically. 
\newline\newline
In the following we list the relevant rate equations for our color sextet model. Notational details and definitions are deferred to appendix \ref{app:Boltzmann}.
Our discussion of Boltzmann equations follows closely the expositions in Refs.\cite{Giudice:2003jh,Strumia:2006qk,Buchmuller:2004nz,Kolb:1979qa} 
including subtleties involving the real intermediate state (RIS) subtraction  
to avoid an overcounting of $2\to2$-scattering contributions. 
In our calculation we work to lowest order and do not take finite temperature effects for propagators and coupling constants into account.
For a discussion on finite temperature effects in the context of thermal leptogenesis, see \cite{Giudice:2003jh}.
\newline
For simplicity, we assume $\lambda/M_2, \tilde\lambda/M_2 \ll g_2, \tilde g_2$ so that the dominant
contributions to baryon number violating $X_2$-decays are given in the top line of figure \ref{fig:X2X1X1Tree}.
The relevant decay rates and scattering cross sections are given by\footnote{Here we give the results in terms of rates where we \textbf{averaged} over spins 
and colors in the initial state and \textbf{summed} over final state quantum numbers. 
However, we express the Boltzmann equations in terms of initial \textbf{and} final state summed rates \cite{Giudice:2003jh}. 
The results for the sextet-gluon scattering rate can also be found in \cite{Chen:2008hh,Manohar:2006ga}.}:
\begin{align}
 \label{eq:GammaX2dd}
 \Gamma(X_2\to dd) 		     = & \frac{M_2  |g_2|^2}{16 \pi}, \\
 \label{eq:GammaX2X1barX1bar}
 \Gamma(X_2\to \bar{X}_1 \bar{X}_1) = & \frac{3 \lambda}{8\pi M_2}\left[\lambda - \tilde{\lambda}\frac{M^2_2 \ \Im \left[ g^{\dagger}_2\tilde{g}_2\right]}{4\pi (M^2_2-\tilde{M}^2_2)}\right], \\
 \Gamma(\bar{X}_2\to X_1 X_1) = & \frac{3 \lambda}{8\pi M_2}\left[\lambda + \tilde{\lambda}\frac{M^2_2 \ \Im \left[g^{\dagger}_2\tilde{g}_2\right]}{4\pi (M^2_2-\tilde{M}^2_2)}\right],
\end{align}
\begin{align*}
 \sigma(gg\to X_2 \bar X_2) = & \frac{\pi \alpha^2_s}{s^2}\frac{2 C(2)_6 d_6}{d^2_8} 	\times			\nonumber \\
	  & \left[\frac{1}{6}\beta [6C(2)_6(4M^2_2+s)+C(2)_8(10M^2_2-s)] \right. 				\nonumber\\
	  & \left.    -\frac{4M^2_2}{s}[C(2)_8 M^2_2 + C(2)_6 (s-2M^2_2)]\ln\frac{1+\beta}{1-\beta}\right], 
\end{align*}
where $\beta = \sqrt{1- \frac{4 M^2_2}{s}}$. The group theory constants are given in tab.~\ref{tab:GroupTheoryConstants}.
\begin{table}[ht!]
\centering
\begin{tabular}{|l|| c | c | c | }
  \hline                        
  $R\hat{=}d_R$ 	& 3 	& 6 	& 8 \\
  $C_R$ 	& 1/2 	& 5/2 	& 3 \\
  $C(2)_R$ 	& 4/3	& 10/3	& 3 \\
  \hline  
\end{tabular}
\caption{\label{tab:GroupTheoryConstants} Group theory constants for $SU(3)$-multiplets. $d_R$ is the dimension of the representation $R$. $C_R$ denotes the Dynkin index $\text{Tr}[T^a_R T^b_R] = C_R \delta^{ab}$, whereas $C(2)_R$ represents the quadratic Casimir $\sum_{a} T^a_RT^a_R = C(2)_R \mathbb{1}_R$.}
\end{table}
\newline
Following our notation of appendix \ref{app:Boltzmann}, we write the Boltzmann equation for $X_2$
\footnote{We neglected all terms of order $\mathcal{O}(\epsilon^2)$ and measure temperatures with respect to $M_2$ by introducing the dimensionless parameter $z=M_2/T$.
We write all rates as thermally averaged quantities $\gamma^{eq}_i$, c.f. App.\ref{app:Boltzmann}.}:
\begin{align}
 \label{eq:BoltzmannX2}
 s H(z) z & \frac{d Y_{X_2}(z)}{dz} =  - \left[X_2 \leftrightarrow \bar X_1 \bar X_1\right] - \left[X_2 \leftrightarrow \bar d \bar d \right] + \left[gg \leftrightarrow X_2 \bar X_2\right] 	\nonumber\\
 = & \left[- \frac{Y_{X_2}}{Y^{eq}_{X_2}} + 1\right] \gamma^{eq}_D + \left[-\frac{Y_{X_2}Y_{\bar X_2}}{Y^{eq}_{X_2}Y^{eq}_{X_2}} +1\right] \gamma^{eq}(gg\to X_2 \bar X_2)   			\nonumber \\
   & - \frac{\bar Y_{X_1}}{Y^{eq}_{X_1}} Br\ \gamma^{eq}_D - \frac{\bar Y_d}{Y^{eq}_d}(1-Br)\gamma^{eq}_D,				    	   
\end{align}
and for $\bar X_2$:
\begin{align}
\label{eq:BoltzmannX2bar}
 s H(z) z & \frac{d Y_{\bar X_2}(z)}{dz} =  - \left[\bar X_2 \leftrightarrow  X_1  X_1\right] - \left[\bar X_2 \leftrightarrow  d  d \right] + \left[gg \leftrightarrow X_2 \bar X_2\right] \nonumber\\
	= &  \left[- \frac{Y_{\bar X_2}}{Y^{eq}_{X_2}} + 1\right] \gamma^{eq}_D + \left[-\frac{Y_{X_2}Y_{\bar X_2}}{Y^{eq}_{X_2}Y^{eq}_{X_2}} +1\right] \gamma^{eq}(gg\to X_2 \bar X_2) 			 \nonumber \\
	&  + \frac{\bar Y_{X_1}}{Y^{eq}_{X_1}} Br\ \gamma^{eq}_D + \frac{\bar Y_d}{Y^{eq}_d}(1-Br)\gamma^{eq}_D.
\end{align}
In eq.(\ref{eq:BoltzmannX2}) we have defined the following quantities:
\begin{align}
  \bar Y_{X_1} \equiv   & Y_{X_1} - Y_{\bar X_1}, \\
  \bar Y_{d}   \equiv   & Y_{d}   - Y_{\bar d},   \\
  \label{eq:epsilon}
  \epsilon \equiv	 & \frac{\gamma^{eq}(X_2 \to \bar X_1 \bar X_1) - \gamma^{eq}(\bar X_2 \to X_1 X_1)}{\gamma^{eq}(X_2 \to \bar X_1 \bar X_1) + \gamma^{eq}(\bar X_2 \to X_1 X_1)},\\
  \label{eq:Br}
  Br \equiv 		 & \frac{\gamma^{eq}(X_2 \to \bar X_1 \bar X_1)}{\gamma^{eq}(X_2 \to \bar X_1 \bar X_1) + \gamma^{eq}(X_2 \to \bar d \bar d)}, \\
  \gamma^{eq}_D \equiv  & \gamma^{eq}(X_2 \to \bar X_1 \bar X_1) + \gamma^{eq}(X_2 \to \bar d \bar d).
 \end{align}
Note that $\bar Y_{d, X_1}$ is zero in thermal equilibrium, so that all source terms in the Boltzmann equation vanish and the particle number for $X_2$ does not change in co-moving coordinates as expected.
Throughout this work, we measure temperatures with respect to $M_2$ and define the dimensionless variable $z = M_2/T$.
Assuming that $X_2$ decays only into $X_1$ and $d$, we have additional relations between the decay rates:
\begin{align}
  \frac{\gamma^{eq}(X_2 \to \bar d \bar d)}{\gamma^{eq}_D} = & 1 - Br ,\\
  \frac{\gamma^{eq}(\bar X_2 \to X_1 X_1)}{\gamma^{eq}_D}  = & \frac{1-\epsilon}{1+\epsilon}Br, \\
  \frac{\gamma^{eq}(\bar X_2 \to dd)}{\gamma^{eq}_D}       = & 1 - \frac{1-\epsilon}{1+\epsilon}Br.
\end{align}
In order not to overestimate washout effects, we need to take into account real intermediate state subtraction (RIS) for the light species \cite{Giudice:2003jh,Strumia:2006qk,Buchmuller:2004nz}. 
The Boltzmann equations for the asymmetry of the light particles $\bar Y_{X_1} = Y_{X_1} - Y_{\bar X_1}$ and $\bar Y_{d} = Y_{d} - Y_{\bar d}$ are expressed in terms of unsubtracted quantities only.
	\begin{align}
	\label{eq:BoltzmannYbarX1unsub} 
	& z H s  \frac{d \bar Y_{X_1}}{dz} =  -2 \left[ 
							\frac{Y_{X_2}-Y_{\bar X_2}}{Y^{eq}_{X_2}} Br \ \gamma^{eq}_D 
						      + \left(\frac{Y_{\bar X_2}}{Y^{eq}_{X_2}} -1 \right) \epsilon\ Br \ \gamma^{eq}_D 
						  \right] 				\nonumber \\
					     & + 4 \left[
							\left(1+\frac{\bar Y_d}{Y^{eq}_d}\right) \gamma^{eq}(dd\to X_1 X_1) 
						      - \left(1+\frac{\bar Y_{X_1}}{Y^{eq}_{X_1}}\right)\gamma^{eq}(X_1 X_1 \to dd) \right. \nonumber \\
						&   \left. - \frac{\bar Y_{X_1}}{Y^{eq}_{X_1}} Br^2 \gamma^{eq}_D
						      - \frac{\bar Y_d}{Y^{eq}_d}(1-Br)Br \ \gamma^{eq}_D
						   \right] 
	\end{align}
	\begin{align}
	\label{eq:BoltzmannYbardunsub}
	& z H s  \frac{d \bar Y_{d}}{dz} =  -2 \left[ 
							\frac{Y_{X_2}-Y_{\bar X_2}}{Y^{eq}_{X_2}} (1-Br) \ \gamma^{eq}_D 
						      - \left(\frac{Y_{\bar X_2}}{Y^{eq}_{X_2}} -1 \right) \epsilon\ Br \ \gamma^{eq}_D 
						  \right] 				\nonumber \\
					     & - 4 \left[
							\left(1+\frac{\bar Y_d}{Y^{eq}_d}\right) \gamma^{eq}(dd\to X_1 X_1) 
						      - \left(1+\frac{\bar Y_{X_1}}{Y^{eq}_{X_1}}\right)\gamma^{eq}(X_1 X_1 \to dd) \right. \nonumber \\
						&   \left. 
						      + \frac{\bar Y_{X_1}}{Y^{eq}_{X_1}} (1-Br)Br \ \gamma^{eq}_D
						      + \frac{\bar Y_d}{Y^{eq}_d}(1-Br)^2 \ \gamma^{eq}_D
						   \right] 
	\end{align}
To obtain the total baryon yield we have to combine the contributions from $X_1$ and $d$,
\begin{align}
 Y_B = \frac{1}{3} \bar Y_d - \frac{2}{3} \bar Y_{X_1}.
\end{align}
Before embarking on a numerical study of the Boltzmann evolution equations for the baryon asymmetry we wish to give a qualitative picture of our model by investigating various 
limits of the parameter space. An example of the relevant reaction rates that enter the Boltzmann equations is shown in fig.~\ref{fig:BoltzmannRates}.
\newline
The quantitative behavior of these rates is well understood and can be found in standard textbooks on cosmology (c.f. e.g. \cite{Kolb:1990vq}).
For the sake of brevity, we will not repeat the discussion here with two exceptions. First, due to the large color charge of $X_2$, one could think that its
strong interactions with gluons keeps $X_2$ in equilibrium so that no baryon asymmetry can be generated. We demonstrate that this is generically not the case. 
And second, we discuss the decay rates of $X_2$ as they play a role in our discussions on $n\bar n$-oscillations later.
\begin{enumerate}
 \item[(a)] Gluon interactions: $\Gamma_{gg\to X_2 \bar X_2 } = n^{eq}_g \langle \text{v} \sigma(gg\to X_2 \bar X_2 )\rangle$; We evaluate the
            thermal average, eq.(\ref{eq:thAv2to2Rate}), numerically and compare it to the Hubble rate, eq.(\ref{eq:HubbleParameter}). 
           This constraint only depends on the strong coupling constant $\alpha_s$ and $M_2$, the mass of $X_2$.
					 The numerical results are summarized in fig.\ref{fig:outofeqRatesContour}.
           Analytically, we can estimate the mass $M^*_2$ for which the two rates are equal at the characteristic
					 temperature $T=M_2\Leftrightarrow z=1$ and find
           $M^*_2 \sim \frac{\alpha^2_s M_{pl}}{\pi^2 g^{1/2}_{*S}}$. Using the RGE-evolved strong coupling constant $\alpha_s$ at the high scale, we find $M^*_2\sim 10^{13-14}$ GeV. 
           Comparing $\Gamma_{gg\to X_2 \bar X_2 }$ and $H$ only at $z=1$, one would conclude that gluon scattering keeps $X_2$ in thermal equilibrium for masses $M_2 < M^*_2$
           because $\left. n^{eq}_g \langle\sigma \text{v}\rangle \right|_{z=1} \sim M_2$ and the Hubble rate falls as $H\sim M^2_2$ at $z=1$. 
           However, this does not take into account the 
           exponential Boltzmann suppression once the temperature falls below the $M_2$-threshold. In that case only gluons in the high energy tail of the distribution have sufficient energy to 
           pair produce the $X_2$. In the reverse process, the rate is proportional to the number density of $X_2$ which also follows a Boltzmann distribution for $T<M_2$. 
           An example of this behavior can be seen in fig.~\ref{fig:BoltzmannRates}. Only for low masses $M_2$ is it possible that the gluon interactions keep $X_2$ in equilibrium 
           in a reasonable temperature range ($1\lesssim z\lesssim 10$) so that the generation of baryon asymmetry is inhibited, c.f. fig.\ref{fig:outofeqRatesContour}. 
 \item[(b)] Decays: $\langle \Gamma(X_2 \to \bar X_1 \bar X_1;\ \bar d \bar d)\rangle$; 
	   For this discussion it is useful to combine the two tree level decay channels of $X_2$, eq.(\ref{eq:GammaX2dd}) and eq.(\ref{eq:GammaX2X1barX1bar}) into the total decay rate. 
	   As discussed before, $\lambda$ is dimensionful and should be of order 
	   $\lambda \sim M_2$ which is why we introduce the dimensionless coupling $\bar\lambda=\lambda/M_2$. Parameterizing the total decay rate in terms of 
	   an effective interaction strength $\Gamma_{X_2} = \frac{1}{4} M_2 \alpha_{\text{eff}}$, where
	   \begin{align}
	   \label{eq:DefAlphaEff}
	   \alpha_{\text{eff}} = \frac{|g_2|^2 + 6 \bar \lambda^2}{4\pi},	      
	   \end{align}
	   we obtain the out of equilibrium relation
	   \begin{align}
	    \alpha_{\text{eff}} 		& < 1.66\times g^{1/2}_{*S} M_2/M_{pl}.
	   \end{align}
\end{enumerate}
The requirement for the decay rates to satisfy the out of equilibrium conditions are not entirely sharp. 
In Ref.\cite{Kolb:1990vq} an approximate analytic analysis of washout effects due to inverse decay processes and $2\to 2$-scattering processes can be found. 
If we introduce the ratio $K = \left.\langle \Gamma_{X_2}\rangle\right|_{z=1}/2 H(1)$ by which the out of equilibrium conditions are violated, one has to include a damping factor for 
the free out of equilibrium result $Y^{\text{free}}_B = 2 \epsilon Br / {g_{*S}}$, so that \cite{Kolb:1990vq}
\begin{align}
\label{eq:YBImproved}
 Y^{imp}_B \approx Y^{\text{free}}_B \times \frac{0.3}{K (\ln K)^{0.6}}, \text{ if $K>1$ }.
\end{align}
For very large $K$, when baryon number violating $2\to2$-scatterings are important, the washout factor decreases the final baryon asymmetry exponentially\cite{Kolb:1990vq}.
We visualize the interesting parameter region for baryogenesis by comparing the Hubble rate $H(z=1)$ with the relevant 
annihilation cross section $n^{eq}_{X_2} \langle v\sigma(gg\to X_2 \bar X_2)\rangle $ as well as 
thermally averaged decay rate\footnote{$K_i(z)$ denote the modified Bessel functions.} $\langle \Gamma_{X_2} \rangle = \frac{K_1(z)}{K_2(z)}\Gamma_{X_2}$ in the $M_2$-$\alpha_{\text{eff}}$-plane 
in fig.~\ref{fig:outofeqRatesContour}.
The main focus in the following paragraphs lies in a numerical treatment of the Boltzmann equations to correctly take washout effects into account beyond the
analytical approximations. This also highlights the fact that not all parameters that give a correct baryon yield in the free out of 
equilibrium approximation, $Y_B = 2 \frac{\epsilon Br}{g_{*S}}$, are viable. In our treatment, we improve previous discussions 
of high scale baryogenesis within the color sextet model found in Refs.~\cite{Arnold:2012sd,Babu:2012vc}.
\subsection{Reducing number of model parameters}
Our model in its general form has a large number of free parameters. In addition to the coupling constants, we have the masses of the $\overset{\text{\tiny{(}} \sim \text{\tiny{)}}}{X}_2$-fields, $\overset{\text{\tiny{(}} \sim \text{\tiny{)}}}{M}_2$ (treating the quarks and $X_1$ massless for the purpose of baryogenesis).
\newline
Exploring the full parameter range of the model including all phases is a formidable task which is beyond the scope of this work. In the one-family limit described in Sec.~\ref{sec:Model}, our model does not contribute new sources to meson- anti-meson mixing. Even if we relaxed this assumption, the large $X_2$-mass helps to satisfy the limits from $K^{0} \bar K^{(0)}$-mixing for example. 
\newline\newline
The relevant parameters for baryogenesis are the $CP$-violating parameter $\epsilon$, eq.(\ref{eq:epsilon}), and the branching fraction $Br$, eq.(\ref{eq:Br}) which reduce to
\footnote{Instead of the phase $\Im\left[{g^{\dagger}_2 \tilde g_2}\right]$ and the mass ratio $\widetilde{M}^2_2/M^2_2$, we mostly work with the effective parameter 
$\epsilon$ that characterizes $CP$-violation.}:
\begin{align}
 \label{eq:epsRED}
 \epsilon 	& = \frac{\tilde{ \lambda}}{ \lambda} \frac{1}{4\pi} \frac{M^2_2 \ \Im\left[{g^{\dagger}_2 \tilde g_2}\right]}{\widetilde{M}^2_2-M^2_2},  
 \\
 \label{eq:BrRED}
 Br 	  	& = \frac{3 \bar \lambda^2}{\frac{1}{2}|g_2|^2 + 3 \bar \lambda^2}.
\end{align}
We combine the phase $\Im\left[{g^{\dagger}_2 \tilde g_2}\right]$, the mass ratio of the heavy sextets and the ratio of the trilinear couplings into a single 
effective $CP$-violating parameter $\epsilon$.
Within this simplified framework, we ask which model parameters give a baryon yield $Y_B$\footnote{The yield $Y_B$ is related to the commonly quoted baryon to photon ration \cite{Bennett:2012zja} $\eta = \frac{n_B - n_{\bar B}}{n_{\gamma}} = (6.19\pm0.14)\times 10^{-10}$.} consistent with experiment \cite{Babu:2012iv}
$ Y_B = \frac{n_B - n_{\bar B}}{s} =  (8.75 \pm 0.23) \times 10^{-11}. $
To estimate the range of viable model parameters, we insert eq.(\ref{eq:epsRED}) and eq.(\ref{eq:BrRED}) into the free out of 
equilibrium decay result which relates the baryon yield $Y_B$ to $\epsilon$ and $Br$,
\begin{align}
 Y^{\text{free}}_B & = 2 \frac{\epsilon Br}{g_{*S}} 
										 = \frac{2}{g_{*S}} \epsilon  \times \frac{3 \bar \lambda^2}{\frac{1}{2} |g_2|^2 +3 \bar \lambda^2},
\end{align}
which is essentially independent of $M_2$\footnote{$M_2$ and $\widetilde M_2$ are expected to be of the same size, 
so that we can parameterize $ R=\widetilde M_2/M_2 $ and the overall scale drops out in $\epsilon$}.
This approximation is valid in parameter regions where all annihilation and decay rates of the baryon number violating $X_2$ are comparably smaller than Hubble $H(z=1)$.
\begin{figure}[!ht]
      \centering
	    \includegraphics[width=0.33\textwidth]{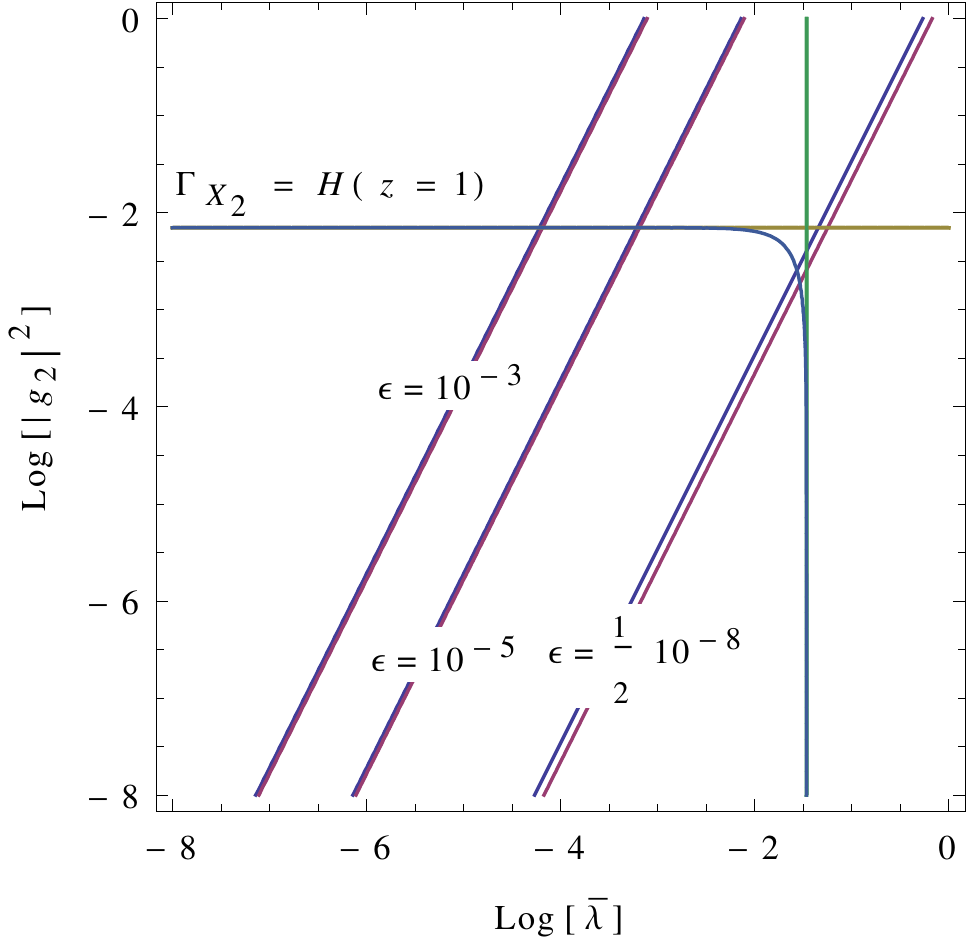}
        \caption{Parameter region consistent with the experimental value $Y_B = (8.75 \pm 0.23) \times 10^{-11}$ using the free out of equilibrium decay estimate. 
        $CP$-violation is characterized by a single parameter, $\epsilon$. The additional contours indicate where the Hubble rate at $z=1$ equals the decay 
        rates $\Gamma_{X_2\to \bar d\bar d}$ (horizontal lines) and $\Gamma_{X_2\to \bar X_1\bar X_1}$ (vertical lines) for $M_2=10^{14}$ GeV.}
        \label{fig:YBFreeDecayEpsilon}
\end{figure}
\subsection{Numerical Results for Baryon Asymmetry}
\label{subsec:NumResBaryogenesis}
\begin{figure}[ht!]
      \centering
	    \includegraphics[width=0.4\textwidth]{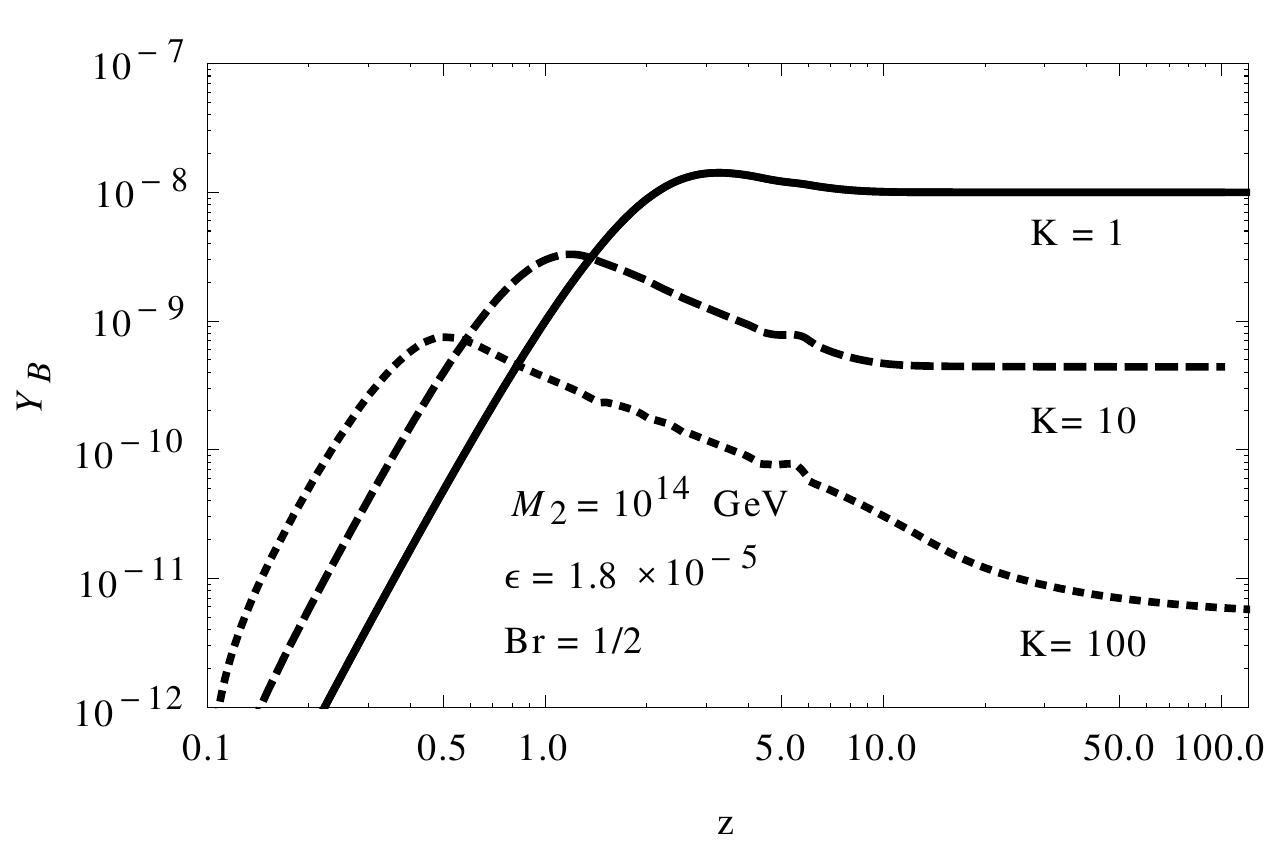}
        \caption{Numerical result of baryon asymmetry $Y_B$ for different $X_2$-decay rates. We adjusted the parameters such that $\epsilon$ and $Br$ remain constant. 
                 Large values of $K = \Gamma_{X_2}/2H(z=1)$ keep $X_2$ in equilibrium longer which leads to washout effects at large $z$.
                }
        \label{fig:YBNumeric}
\end{figure}

Following the general considerations and the analytical estimates of parameter regions, where baryogenesis is feasible, we show numerical results for the coupled system of 
Boltzmann equations where a free decay scenario is not applicable. 
We verified that our numerical results agree with analytic estimates in the free out of equilibrium decay regime of the model (large $M_2$, small $\alpha_{\text{eff}}$). 
\newline
In order to investigate the influence of the decay rate of $X_2$ relative to the Hubble rate, we increase the cubic scalar coupling $\lambda$ together with $g_2$ 
gradually so that the free out of equilibrium conditions are not satisfied any more. To make the results comparable, we keep $\epsilon$ and $Br$ constant.
Doing so, the free out of equilibrium calculation tells us that the baryon yield $Y_B = 2 \epsilon Br /g_{*S}$ remains constant as well.
We also compare our numerical results to the improved analytical solution given in eq.(\ref{eq:YBImproved}) if applicable. 
Starting with a parameter point, where the coupling constants are chosen so that the decay rates are equal to the Hubble rate,
\begin{align}
  M_2 & = 10^{14}\text{GeV}, 			 							 && \widetilde M_2 =  2 \ M_2, 						\nonumber \\
  \bar{\lambda}^2 & = \frac{8\pi}{3} 1.66\ g^{1/2}_{*}\frac{M_2}{M_{pl}} = (0.034)^2, && \tilde \lambda = \lambda,						\nonumber \\
  |g_2|^2 & = 16 \pi\ 1.66\ g^{1/2}_{*} \frac{M_2}{M_{pl}} = 0.007,  		 &&  \Im\left[g^{\dagger}_2 \tilde g_2\right] = 0.1  |g_2|^2, 	\nonumber \\
  \epsilon = & 1.8\times 10^{-5},				 					 && Br = 0.5,								\nonumber
\end{align}
and increasing $\alpha_{\text{eff}}$, eq.(\ref{eq:DefAlphaEff}), by factors of 10 corresponds to a violation of one of Sakharov's conditions \cite{Sakharov:1967dj}. 
As expected, for large $K=\langle\Gamma_{X_2}\rangle/2H(1)$, the net baryon asymmetry decreases considerably.
For fig.\ref{fig:YBNumeric}, we keep $\epsilon$, $Br$ and $M_2$ fixed, so that $ Y^{\text{free}}_B = \frac{2 \epsilon Br}{g_{*S}} \approx 1.8 \times 10^{-7}$ is constant.
\begin{table}[h!]
\centering
\begin{tabular}{|c|| c |c | c| }
  \hline                        
  $K$ 			& $Y^{\text{free}}_B$		& $Y^{\text{imp}}_B$		& $Y^{\text{num}}_B$ 		\\
  \hline\hline
  $1$ 			& $1.8 \times 10^{-7}$ 		& $-$				& $9.9\times 10^{-9}$		\\
  $10$ 			& $1.8 \times 10^{-7}$		& $3.2\times 10^{-9}$		& $4.3 \times 10^{-10}$		\\
  $100$ 		& $1.8 \times 10^{-7}$		& $9.7\times{10^{-11}}$		& $5.9 \times 10^{-12}$		\\
  \hline  
\end{tabular}
\caption{\label{tab:YBCompResults} Comparison between free out of equilibrium ($Y^{\text{free}}_B$), 
	  washout improved analytical ($Y^{\text{imp}}_B$, eq.(\ref{eq:YBImproved})) and
	  numerical ($Y^{\text{num}}_B$) results for the baryon asymmetry.}
\end{table}
\newline
Note that the results in tab.\ref{tab:YBCompResults} show that our numerical results can vary up to a factor of ten from the washout corrected analytic results $Y^{\text{imp}}_B$, eq.(\ref{eq:YBImproved}).
We also increased the $CP$-violating parameter $\epsilon$ and found roughly linear dependence. The same is true when the branching fraction $Br$ is varied.
In fact, our model reproduces all aspects of the high scale toy model for baryogenesis discussed in Ref.~\cite{Kolb:1979qa}.
\begin{figure}[ht!]
      \centering
	    \includegraphics[width=0.4\textwidth]{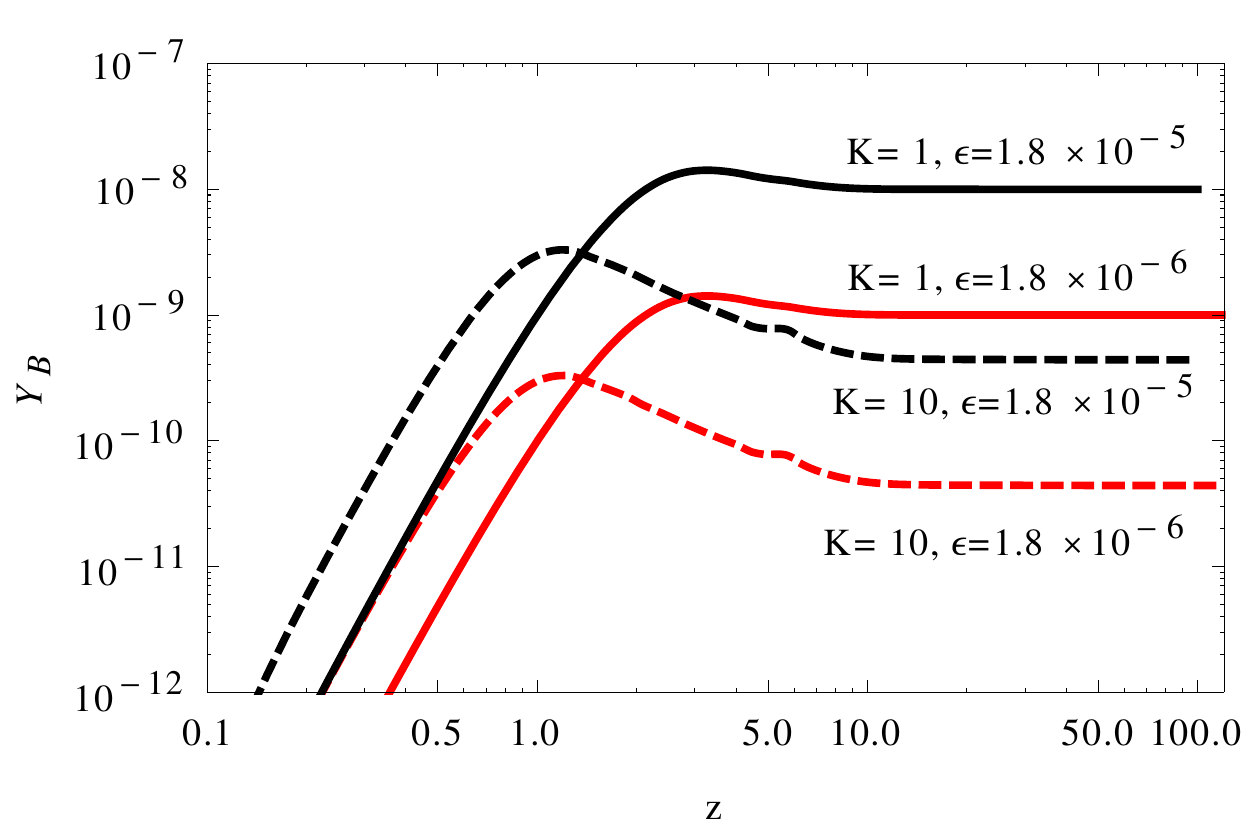}
        \caption{Comparing numerical results for the baryon asymmetry for different $K$ and $\epsilon$, keeping $M_2=10^{14}$ GeV fixed. 
		 We find excellent agreement with the expectations of Ref.~\cite{Kolb:1979qa}. }
        \label{fig:YBNumericEps}
\end{figure}
%
%
\section{$\mathbf{n\bar n}$-oscillations}
\label{sec:NNbarOscillations}
\begin{figure}[!ht]
\centering
	    \includegraphics[width=0.15\textwidth]{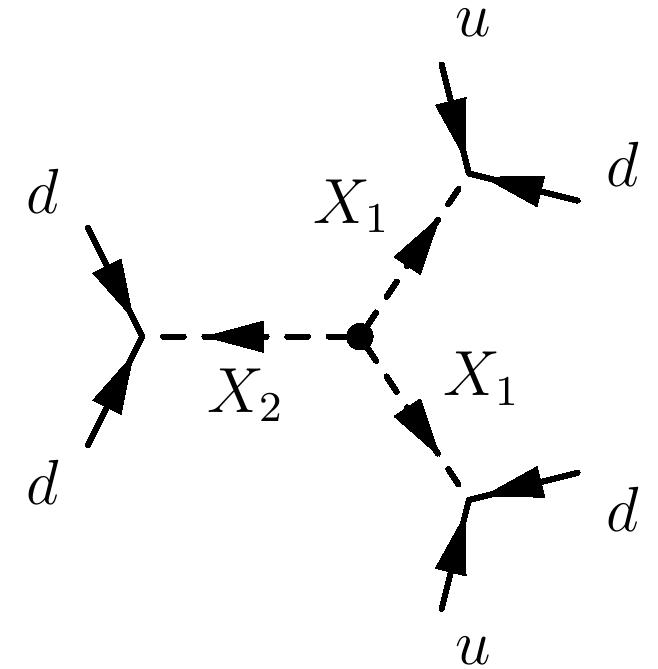}
            \caption{Contribution to $n\bar n$-oscillations via $\Delta B = 2 $ process involving the scalar sextet fields.}
            \label{fig:nnbarOscFeynmanDiag}
\end{figure}
\begin{figure}[!ht]
      \centering
	    \includegraphics[width=0.35\textwidth]{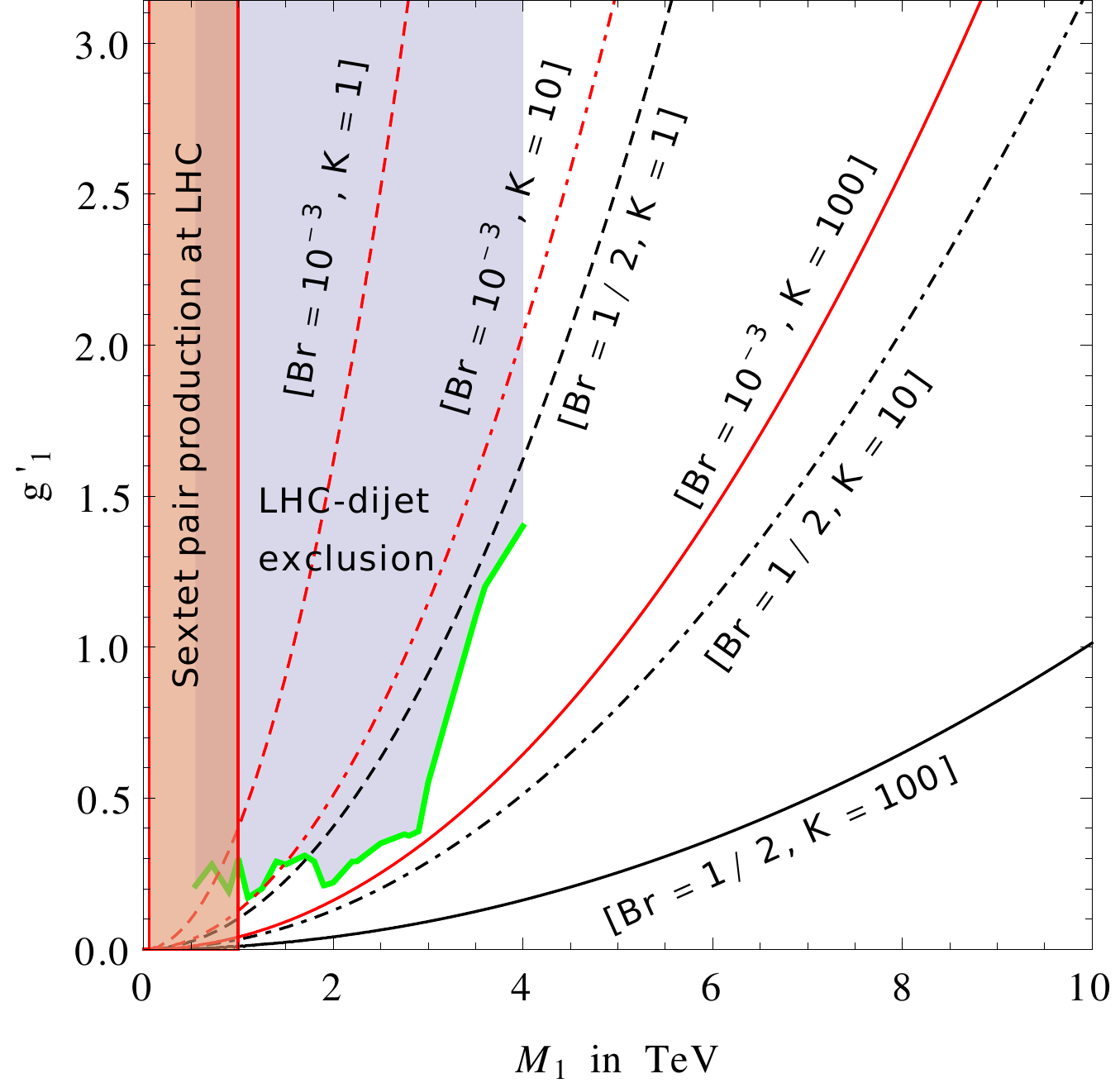}
        \caption{Contourplot of $\Delta m$ in the $g'_1-M_1$-plane for different values of $K = \Gamma_{X_2}/2 H(1)$ and branching fraction $Br$ of $X_2$. 
                 Different contours represent values where $\Delta m$ as written in eq.(\ref{eq:DeltaM}) is equal to the expected 
                 sensitivity of future $n\bar n$-oscillation experiments, $\Delta m^{\text{fut}} = 10^{-35}$ GeV.
                 Note that current limits on $n \bar n$-oscillations can easily be avoided if the decay rate $\Gamma_{X_2}$ is small, there is no obstruction from baryogenesis.
                 If we see $n\bar n$-oscillations in near future experiments, the allowed parameter region is to the upper left of the contours. 
                 We also include collider limits on color sextets from LHC-dijet searches that have been analyzed in Ref.~\cite{Richardson:2011df}.
                 The orange countour represents the simulated limits for color sextet pair production at the 14 TeV LHC with $\mathcal{L}=100 fb^{-1}$ integrated luminosity.
                 }
        \label{fig:NBarNOscillationProspects}
\end{figure}

Neutron anti-neutron oscillation experiments can directly probe the structure of baryon number violation \cite{Kamyshkov:2002vm,Mohapatra:2009wp}. 
In comparison to proton decay experiments which probe $\Delta B = 1, \Delta L\ \text{odd}$-modes (e.g. $p\to e^+ \pi^0$), 
$n\bar n$-oscillations are intimately related to $\Delta B=2$ processes. Future experiments such as the European Spallation Source (ESS)
could push the bounds on this mode of matter instability to $\tau \approx 10^{36}$ years\footnote{\url{http://europeanspallationsource.se/fundamental-and-particle-physics}} exceeding 
current limits on the proton lifetime $\tau_p > 8.2\times10^{33}$ years \cite{Nishino:2009aa}.
A number of models that predict $n\bar n$-oscillations have been considered in the past, c.f. e.g.~\cite{Kuzmin:1970nx,Chacko:1998td,Babu:2012vc,Arnold:2012sd,Dutta:2005af}.
\newline
The fact that $n\bar n$-oscillations require baryon number violation suggests that the underlying dynamics of these 
low energy processes can be linked to baryogenesis as pointed out by Kuzmin \cite{Kuzmin:1970nx} and others. 
In the presence of a TeV-scale colored scalar, 
$n\bar n$-oscillation experiments can probe energy scales up to $10^{15\text{-}16}$ GeV \cite{Arnold:2012sd}, which we take to be the scale of baryon number violating interactions via $X_2$-decays.
Light color sextet scalars, such as $X_1$ ($M_1 \sim $few TeV), can be present in non-supersymmetric $SO(10)$-models that lead to gauge coupling unification as described in detail in \cite{Babu:2012vc}. 
\newline
In the previous section we focused on a careful treatment of baryogenesis by solving Boltzmann equations numerically.
Here we are going to relate these results to the discovery potential of $n\bar n$-oscillations in present and future experiments.
\newline\newline
Neutron-anti-neutron oscillation experiments are sensitive to the transition matrix element $\Delta m = \langle \bar n| \mathcal{H}_{\text{eff}}|n\rangle $. 
The effective low energy Hamiltonian $\mathcal{H}_{\text{eff}}$ for $n\bar n$-transitions in our model is given in Ref.~\cite{Arnold:2012sd}\footnote{A similar result was obtained in \cite{Babu:2012vc}}. Neglecting 
the Yukawa-coupling to left-handed quarks, $g_1$, $\Delta m$ is characterized by \cite{Arnold:2012sd}
\begin{align}
\label{eq:nnbaroscillation}
 \Delta m = 2 \lambda \beta^2 \frac{\left|(g'_1)^2 g_2\right|}{3 M^4_1 M^2_2},
\end{align}
where $\beta = 0.01 \text{GeV}^3$ has been determined in lattice gauge theory \cite{Tsutsui:2004qc}. 
The current limit on $n\bar n$-oscillations is given by \cite{Abe:2011ky} $\Delta m <2 \times 10^{-33}$ GeV and could be improved by a few orders 
of magnitude in suggested future experiments\footnote{In \cite{Babu:2013yww}, four orders of magnitude improvements on the free oscillation probability (corresponds to two orders of magnitude in $\Delta m$) are estimated for a 
1MW spallation target at Project X. The limits for ESS should be at least comparable.}.
\newline
\newline
If we restrict ourselves to the one-family scenario, we can express $\Delta m$ in terms of the decay rates by substituting $\lambda$ and $g_2$.
\begin{align}
 \Delta m = 2 \beta^2 \sqrt{\frac{128\pi^2}{3}} \sqrt{Br(1-Br)} \frac{|g'_1|^2}{3 M^2_2 M^4_1}\ \Gamma_{X_2}.
\end{align}
For successful baryogenesis, the $X_2$s have to decay out of equilibrium. 
This is only possible if the $K=\Gamma_{X_2}/2H(z=1)$-factor as defined in the previous section is not 
larger than $\mathcal{O}(100)$ for a $CP$-violating parameter $\epsilon\approx10^{-5}$.
Taking this into account, we can relate successful baryogenesis to $n\bar n$-oscillations by setting $\Gamma_{X_2} < K\times H(1) = K \times 1.66\ g^{1/2}_{*S} M^2_2/M_{pl}$.
With this substitution $M_2$ drops out of $\Delta m$, and we get an upper bound on the transition rate:
\begin{align}
\label{eq:DeltaM}
 \Delta m < \frac{2 \beta^2 \times 1.66 g^{1/2}_{*S}}{3 M_{pl}}\sqrt{\frac{128 \pi^2}{3}} \sqrt{Br(1-Br)} \frac{|g'_1|^2}{M^4_1}K.
\end{align}
Written in this form, $\Delta m$ is tied to $Br$ and $K$ which directly enter our baryogenesis analysis described above.
One is left with parameters that concern the light sextet $X_1$. This allows us to consistently combine $n\bar n$-oscillations, baryogenesis and LHC-phenomenology for our color sextet model. 
For $M_1$ around the TeV-scale one can search for signals of this particle at the LHC. 
In principle, we might be able to measure its coupling $g'_1$ to $u_R,\ d_R$-quarks picking out one particular point in fig.\ref{fig:NBarNOscillationProspects}. 
We comment on possible collider signals in the next section.
\newline
Note that there is some tension between demanding visible $n\bar n$-oscillations in the future, successful baryogenesis and collider constraints. 
Even for $Br=1/2$, where $\sqrt{Br(1-Br)}$ is maximized, a $K$-factor of 1 is almost completely excluded. 
Increasing $K$ to larger values, it becomes important to take washout effects of the baryon asymmetry into account
and one is forced to go beyond the free out of equilibrium decay scenario used in \cite{Arnold:2012sd}.
%
%
\section{Constraints from Neutron Electric Dipole Moments and Colliders}
\label{sec:EDMLHC}
%
%
\begin{figure}[!ht]
\centering
	    \includegraphics[width=0.25\textwidth]{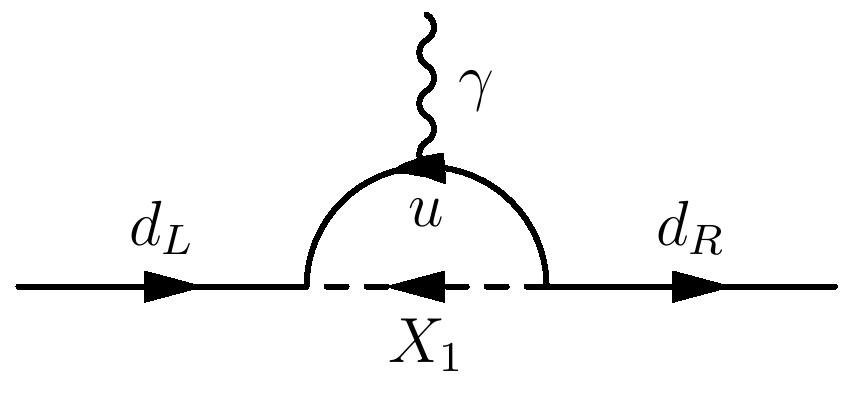}
            \caption{Down-quark contribution to the electric dipole moment of the neutron.}
            \label{fig:DownEDM}
\end{figure}
Since our model includes additional sources of $CP$-violation, in principle we have to worry about contributions to quark electric dipole moments (EDMs) coming from 
diagrams such as fig.~\ref{fig:DownEDM}. 
Using $SU(6)$-wavefunctions, the quark EDMs are related to the dipole moment of the neutron $d_N$ by, 
$
 d_N = \frac{4}{3} d_d - \frac{1}{3} d_u \simeq \frac{4}{3} d_d
$\cite{Czarnecki:1997bu}.
In the Standard Model, quark EDMs are first generated at 3-loop \cite{Czarnecki:1997bu} and are
additionally suppressed by the $GIM$-mechanism leading to a prediction for the 
neutron-EDM of\footnote{Taking long distance effects of a six quark operator into account, Ref.\cite{Mannel:2012qk} finds $d_N = \mathcal{O}(10^{-31})$ e cm. } $d_N = \mathcal{O}(10^{-34})$ e cm. 
The current experimental bound is given in \cite{Serebrov:2013tba}, $d^{\text{exp}}_N < 5.5 \times 10^{-26}\ \text{e cm}$. 
\newline
The tiny Standard Model contributions allow us to estimate limits for our model parameters. 
Specifically, we obtain restrictions on $g'_1$ from $X_1$ contributions to the down-quark-EDM via the diagram shown in fig.~\ref{fig:DownEDM}. 
For a nonzero $g_1$ (so far we used $g_1=0$) in the one generation approximation, the leading contribution of $X_1$ to the electric dipole moment of the down-quark is given by\cite{Arnold:2012sd}:
\begin{align}
 |d_d| \simeq \frac{m_u}{6 \pi^2 M^2_1}\log\left(\frac{M^2_1}{m^2_u}\right) \left|\Im\left[g^{ }_1(g'^{ }_1)^{*})\right]\right|\text{ e cm}.
\end{align}
If we use the QCD scale $\Lambda_{\text{QCD}}\approx 1$ GeV instead of $m_u$ we estimate,
\begin{align}
\label{eq:EDMconst}
 \left|\Im\left[g_1(g'_1)^{*})\right]\right|\lesssim 8\times 10^{-3},
\end{align}
for $M_1 = 2 \text{ TeV}$.
Assuming that $g_1$ is comparable to $g'_1$ and there are no small phases in this sector, eq.(\ref{eq:EDMconst}) suggests, that $g'_1 = \mathcal{O}(0.1)$. 
Comparing this to the parameter region preferred by baryogenesis and visible $n \bar n$-oscillations points towards a light $M_1$, c.f. fig.~\ref{fig:NBarNOscillationProspects}. 
%
%
\newline
\newline
%
%
The production cross sections and decay rates of color sextet scalars have been studied intensively 
at tree and loop level, c.f. e.g.\cite{Mohapatra:2007af,Han:2009ya,Chen:2008hh,Berger:2010fy,Zhan:2013sza} and references therein. 
Depending on the range of couplings and mass of the sextet, either single- or pair-production can be dominant.
Most of the work focuses on scalar sextets with Standard Model quantum numbers $\Phi\sim(6,1,4/3)$ which 
leads to same-sign top pair production via an s-channel exchange\cite{Mohapatra:2007af}.
However, a majority of the results such as production cross sections or decay rates for these scalars can be extended to different quantum numbers \cite{Berger:2010fy}. 
We are interested in the LHC-phenomenology of $X_1 \sim (\bar6,3,-1/3)$ which couples to $u_R d_R$ rather than $u_R u_R$. 
In terms of single scalar production, both cases should be comparable as the suppression of the down-quark parton distribution function is compensated by the combinatorics of the $u,\ d$ initial state\cite{Han:2009ya}.
In our model with $X_1$ in the $(\bar 6, 1, -1/3)$-representation, the expected signals entail a resonance in the invariant dijet mass distribution (single production) 
or two equal mass dijets (pair production)\cite{Richardson:2011df}.
Note that the authors of Ref.~\cite{Richardson:2011df} obtained limits on the diquark mass of 1 TeV by simulating $\mathcal{L} = 100 \text{ fb}^{-1}$ integrated luminosity for the 14 TeV LHC 
by considering the pair production channel from gluons. This analysis is independent of the unknown sextet Yukawa-couplings and gives the pair production limit in fig.\ref{fig:NBarNOscillationProspects}.
\newline
More interesting for us at this point is the pure dijet analysis (single production) for a sextet 
in the $(6,1,4/3)_{\text{SM}}$-representation in the same article\footnote{\cite{Zhan:2013sza} use higher statistics dijet data from 
ATLAS and CMS to set similar limits on what we call $g_1$ and $g'_1$ but only consider masses up to 2 TeV.}. 
As a final result they show exclusion plots for the coupling of the scalar sextet to right-handed quarks ($u_Ru_R$) depending on the mass of the sextet. 
Since no special features of the $u_Ru_R$ channel such as same sign tops were used in their analysis, we convert their limits to our $g'_1$-$M_1$-plane.
In fig.~\ref{fig:NBarNOscillationProspects} we compared the collider limits with the estimates of $n\bar n$-oscillations. 
To reiterate our conclusion from before; In parameter regions that are not excluded by LHC-searches it is possible to achieve visible $n\bar n$-oscillations and successful baryogenesis. However, washout effects are important and a simple out of equilibrium calculation for the baryon asymmetry is not applicable.
\section{Conclusion}
We studied a minimal model that leads to baryon number violation at tree level but no proton decay. This theory has been suggested in \cite{Arnold:2012sd} and a similar field content was studied in connection to non-supersymmetric gauge coupling unification in \cite{Babu:2012vc}.
The model contains a novel color sextet scalar at the TeV-scale ($X_1 \sim (\bar6,1,-1/3)$) as well as 
two additional scalar sextets at a scale of $\mathcal{O}(10^{14-15})$ TeV ($X_2,\ \widetilde X_2 \sim (\bar6,1,2/3)$). 
In this setup, oscillations between neutrons and anti-neutrons are one of the primary signals of baryon number violating physics.
Since baryon number violation is also one of the key ingredients to explain baryogenesis, it is natural to investigate potential connections between these two phenomena. 
With this motivation and the fact that the sextet scalars have large color charges, we studied a system of coupled Boltzmann equations to accurately predict the evolution 
of the baryon asymmetry in the early Universe in regimes where simple analytic results are not available. 
We were able to relate $n \bar n$-oscillations to baryogenesis in such a way that only two high scale parameters $Br$, the branching ratio of the decay $X_2 \to \bar X_1 \bar X_1$ and 
the out of equilibrium factor $K = \Gamma_{X_2}/2H(1)$ enter the analysis. Taking into account collider constraints on the light sextet $X_1$ we demonstrated that, if we see $n\bar n$-oscillations
at future experiments, such as the suggested European Spallation Source (ESS), washout effects for baryogenesis become important. This is summarized in fig.\ref{fig:NBarNOscillationProspects}.

\section{Acknowledgement}
We thank Mark Wise for detailed discussions throughout the project and careful reading of the manuscript. 
We also acknowledge useful comments by Pavel Fileviez Perez, David Sanford and Yue Zhang.
This work is supported by the DOE Grant \# DE-SC0011632.
\appendix
\section{\label{app:Boltzmann} Details of Boltzmann equations}
In this appendix, we collect a number of well known formulas relevant for our Boltzmann equations (\ref{eq:BoltzmannX2}, \ref{eq:BoltzmannX2bar},\ref{eq:BoltzmannYbarX1unsub},\ref{eq:BoltzmannYbardunsub}). Most of this material is a summary of Refs.~\cite{Kolb:1979qa,Giudice:2003jh,Strumia:2006qk,Buchmuller:2004nz}.
\newline 
In the presence of fast elastic scattering events, the particle species are kept in \textit{kinetic} equilibrium. In this regime, the phase space distribution functions can be approximated by a Maxwell-Boltzmann distribution. Each particle species is then characterized by its chemical potential $\mu_i$ or by its total abundance $n_i$ respectively. These quantities can only be altered by inelastic processes. If these occur fast enough, \textit{chemical} equilibrium will be maintained and the number- and energy-densities follow their equilibrium values $n^{eq}_i$ and $\rho^{eq}_i$ respectively.
The phase space integral can be evaluated exactly. For Maxwell-Boltzmann statistics, one finds \cite{Kolb:1979qa}\footnote{Results for Fermi-Dirac and 
Bose-Einstein statistics can be found in the appendix of \cite{Kolb:1979qa}.}:
\begin{align}
 n^{eq}_{MB} = & g_i \int \frac{d^3p}{(2\pi)^3} f^{eq}(p)  =  \int \frac{d^3p}{(2\pi)^3} e^{-(E_i - \mu_i)/T}	 \nonumber \\
	     = & g_i \int \frac{d^3p}{(2\pi)^3} \exp \left[-(\sqrt{\vec{p}^2 + m^2_i} - \mu_i)/T\right] 		 \nonumber \\
	     = & g_i \frac{T^3}{2\pi^2}e^{-\mu_i/T}z^2_i K_2(z_i),							 \nonumber
\end{align}
where $z_i=m_i/T$. The high- and low-temperature expansion is:
\begin{align}
 n^{eq}_{MB} = 
	      \begin{dcases}
		      g_i \left(\frac{mT}{2\pi}\right)^{3/2} e^{-(m-\mu)/T}\left[1+\frac{15}{8z} + \cdots \right], & m \gg T \\
		      g_i \frac{T^3}{\pi^2}e^{\mu/T}\left[1-\frac{1}{4}z^2 + \cdots \right],               	  & m \ll T 
	      \end{dcases}     
\end{align}
$g_i$ counts the number of internal degrees of freedom of a given species (e.g. $g_{\gamma}=2, \ g_{\text{glue}}=8\times2=16$, etc). In \textit{kinetic} equilibrium the phase space density is 
conveniently parameterized as
\begin{align}
					f(p) = f^{eq}(p) \frac{n}{n^{eq}}.
\end{align}
Following the notation of \cite{Giudice:2003jh,Strumia:2006qk}, the Boltzmann equation describing the time evolution of abundance $n_X$ of some species $X$ is:
  \begin{align}
					\label{eq:BoltzmannTimeParameter}
					\dot n_X + 3 H n_X = - \sum_{a,i,j} \Delta_s \left[X\ a \cdots \leftrightarrow i\ j \cdots \right],
  \end{align}
where the Hubble rate written in terms of $z$:
\begin{align}
			\label{eq:HubbleParameter}
			 H = \frac{\dot a}{a} = \sqrt{\frac{8 \pi \rho }{3 m_{pl}}} 
														= 1.66 g^{1/2}_{*S} \frac{T^2}{m_{pl}} 
														= 1.66 g^{1/2}_{*S} \frac{m^2_X}{m_{pl}} \frac{1}{z^2}.
\end{align}
$\Delta_s$ denotes a symmetry factor which depends on the number of $X$ particles that are created or destroyed in a particular reaction, e.g. $\Delta_s = 2$ for a reaction $XX \to ij$.
\newline
In the definition of the Hubble rate, we encounter the effective number of degrees of freedom $g_{*S}$ that contribute to the energy density $\rho$ of the Universe. Since the relevant epoch is radiation dominated, it counts the relativistic degrees of freedom at a given temperature \cite{Kolb:1990vq}. 
The Standard Model field content leads to $g_{*S}=106.75$ above the electroweak phase transition, but gets modified 
by new relativistic species that are in thermal equilibrium (in our case this would be the light $X_1$-field for example). 
\newline
In eq.(\ref{eq:BoltzmannTimeParameter}) we defined
\begin{align}
 \left[X\ a \cdots \leftrightarrow i\ j \cdots \right] \equiv & \frac{n_X n_a \cdots}{n^{eq}_X n^{eq}_a \cdots} \gamma_{eq}(X\ a \cdots \rightarrow i\ j \cdots)  \nonumber \\
							      - & \frac{n_i n_j \cdots}{n^{eq}_i n^{eq}_j \cdots} \gamma_{eq}(i\ j \cdots \rightarrow X\ a \cdots). 
\end{align}
The $\gamma_{eq}(a b \cdots \rightarrow c d \cdots)$ denote thermally averaged reaction rates for a given process $(a b \cdots \rightarrow c d \cdots)$. 
In the context of Boltzmann equations for baryogenesis, one often encounters the special cases of particle decay and $2\rightarrow 2$-scattering. The results for these processes are given by \cite{Giudice:2003jh,Strumia:2006qk,Buchmuller:2004nz}\footnote{$\gamma_{eq}(X\rightarrow ij)$ is related to the thermally averaged decay rate $\langle\Gamma(X\to ij)\rangle$ via $\gamma_{eq}(X\rightarrow ij)= n^{eq}_X \langle\Gamma(X\to ij)\rangle$.}:
\begin{align}
 \gamma_{eq}(X\rightarrow ij) = & \int \frac{d^3p_X}{(2\pi)^3 2E_X} f^{eq}_X(p_X) \frac{d^3p_i}{(2\pi)^3 2E_i} 
																			 \frac{d^3p_j}{(2\pi)^3 2E_j} \times \nonumber \\
																& (2\pi)^4 \delta^{(4)}(p_X-p_i-p_j) \left|M(X\to ij)\right|^2 \nonumber\\
															= & n^{eq}_X \frac{K_1(z)}{K_2(z)} \Gamma_X,
\end{align}
where $\Gamma_X$ is the decay width in the rest system of the particle summed over all initial \textbf{and} final state spins and colors. $K_1(z)$ and $K_2(z)$ denote the Bessel $K$-functions. 
Introducing the Lorentz-invariant measure $\widetilde{d^3p} \equiv \frac{d^3p}{(2\pi)^3 2E_p}$, the thermally averaged 2-body scattering rate is given by\footnote{The second step is valid if the cross section only depends on $s$ and not on the thermal motion relative to the background plasma \cite{Giudice:2003jh}. $\gamma^{eq}$ is related to the thermally averaged cross section via $\langle \text{v} \sigma(ab\to X) \rangle = n^{eq}_a n^{eq}_b \gamma^{eq}(ab\to X)$.\newline}:
 \begin{align}
 \label{eq:thAv2to2Rate}
  \gamma_{eq} (X a \rightarrow ij) = & \int \widetilde{d^3p_X} \int \widetilde{d^3p_a} f^{eq}_X(p_X) f^{eq}_a(p_a) \int \widetilde{d^3p_i} \int\widetilde{d^3p_j} \times \nonumber \\
				      & (2\pi)^4\delta^{(4)}(p_X+p_a-p_i-p_j) \left|M(X a \to ij)\right|^2 \nonumber\\
				    = & \frac{T}{32 \pi^4}\int^{\infty}_{s_{min}} ds\ s^{3/2} \lambda(1,M^2_x/s,M^2_a/s) \sigma(s) K_1\left(\frac{\sqrt{s}}{T}\right).
 \end{align}
$\sigma(s)$ is the total cross section summed over all initial \textbf{and} final state spins and colors. $\lambda(a,b,c) = (a-b-c)^2-4bc$ is the usual kinematic function. 
$s$ denotes the Mandelstam variable and $s_{min} = \text{max}[(m_X+m_a)^2,(m_i+m_j)^2]$ picks out the threshold energy for a given process. 
\newline
\newline
In order to absorb the dilution of particle species due to the expansion of the Universe reflected in the Hubble term $ 3 H n_X$ (cf. eq.(\ref{eq:BoltzmannTimeParameter})), it is convenient 
to introduce comoving coordinates, $\ Y_i = n_i/s$, where the entropy density of the Universe, $s$, can be expressed as:
\begin{align}
\label{eq:entropyDensity}
  s = & \frac{2 \pi^2}{45} g_{*S}(T) T^3 = \frac{2 \pi^2}{45} g_{*S}(T) \frac{m^3_X}{z^3},
 \end{align}
with the \textit{effective number of degrees of freedom in entropy}
\begin{align}
  g_{*S}(T) = & \sum_{\text{i=bosons}}g_i \left(\frac{T_i}{T}\right)^3 
							+ \frac{7}{8} \sum_{\text{i=fermions}}g_i \left(\frac{T_i}{T}\right)^3.
\end{align}
When \textbf{all} relativistic species are in equilibrium at the same temperature one obtains $g_{*S}(T)=g_{*}$. 
Furthermore, we prefer to write the Boltzmann equations in terms of $z$ instead of time $t$.
In these variables the Boltzmann equation (\ref{eq:BoltzmannTimeParameter}) becomes:
\begin{align}
 \label{eq:BoltzmannZ}
 z H(z) s(z) \frac{dY_X}{dz} = - \sum_{a,i,j} \Delta_s \left[X \ a \cdots \leftrightarrow i \ j \cdots \right].
\end{align}
%
\bibliographystyle{apsrev4-1}
\bibliography{Literatur}

\begin{thebibliography}{42}%
\makeatletter
\providecommand \@ifxundefined [1]{%
 \@ifx{#1\undefined}
}%
\providecommand \@ifnum [1]{%
 \ifnum #1\expandafter \@firstoftwo
 \else \expandafter \@secondoftwo
 \fi
}%
\providecommand \@ifx [1]{%
 \ifx #1\expandafter \@firstoftwo
 \else \expandafter \@secondoftwo
 \fi
}%
\providecommand \natexlab [1]{#1}%
\providecommand \enquote  [1]{``#1''}%
\providecommand \bibnamefont  [1]{#1}%
\providecommand \bibfnamefont [1]{#1}%
\providecommand \citenamefont [1]{#1}%
\providecommand \href@noop [0]{\@secondoftwo}%
\providecommand \href [0]{\begingroup \@sanitize@url \@href}%
\providecommand \@href[1]{\@@startlink{#1}\@@href}%
\providecommand \@@href[1]{\endgroup#1\@@endlink}%
\providecommand \@sanitize@url [0]{\catcode `\\12\catcode `\$12\catcode
  `\&12\catcode `\#12\catcode `\^12\catcode `\_12\catcode `\%12\relax}%
\providecommand \@@startlink[1]{}%
\providecommand \@@endlink[0]{}%
\providecommand \url  [0]{\begingroup\@sanitize@url \@url }%
\providecommand \@url [1]{\endgroup\@href {#1}{\urlprefix }}%
\providecommand \urlprefix  [0]{URL }%
\providecommand \Eprint [0]{\href }%
\providecommand \doibase [0]{http://dx.doi.org/}%
\providecommand \selectlanguage [0]{\@gobble}%
\providecommand \bibinfo  [0]{\@secondoftwo}%
\providecommand \bibfield  [0]{\@secondoftwo}%
\providecommand \translation [1]{[#1]}%
\providecommand \BibitemOpen [0]{}%
\providecommand \bibitemStop [0]{}%
\providecommand \bibitemNoStop [0]{.\EOS\space}%
\providecommand \EOS [0]{\spacefactor3000\relax}%
\providecommand \BibitemShut  [1]{\csname bibitem#1\endcsname}%
\let\auto@bib@innerbib\@empty
\bibitem [{\citenamefont {Cohen}\ \emph {et~al.}(1993)\citenamefont {Cohen},
  \citenamefont {Kaplan},\ and\ \citenamefont {Nelson}}]{Cohen:1993nk}%
  \BibitemOpen
  \bibfield  {author} {\bibinfo {author} {\bibfnamefont {A.~G.}\ \bibnamefont
  {Cohen}}, \bibinfo {author} {\bibfnamefont {D.}~\bibnamefont {Kaplan}}, \
  and\ \bibinfo {author} {\bibfnamefont {A.}~\bibnamefont {Nelson}},\ }\href
  {\doibase 10.1146/annurev.ns.43.120193.000331} {\bibfield  {journal}
  {\bibinfo  {journal} {Ann.Rev.Nucl.Part.Sci.}\ }\textbf {\bibinfo {volume}
  {43}},\ \bibinfo {pages} {27} (\bibinfo {year} {1993})},\ \Eprint
  {http://arxiv.org/abs/hep-ph/9302210} {arXiv:hep-ph/9302210 [hep-ph]}
  \BibitemShut {NoStop}%
\bibitem [{\citenamefont {Luty}(1992)}]{Luty:1992un}%
  \BibitemOpen
  \bibfield  {author} {\bibinfo {author} {\bibfnamefont {M.}~\bibnamefont
  {Luty}},\ }\href {\doibase 10.1103/PhysRevD.45.455} {\bibfield  {journal}
  {\bibinfo  {journal} {Phys.Rev.}\ }\textbf {\bibinfo {volume} {D45}},\
  \bibinfo {pages} {455} (\bibinfo {year} {1992})}\BibitemShut {NoStop}%
\bibitem [{\citenamefont {Affleck}\ and\ \citenamefont
  {Dine}(1985)}]{Affleck:1984fy}%
  \BibitemOpen
  \bibfield  {author} {\bibinfo {author} {\bibfnamefont {I.}~\bibnamefont
  {Affleck}}\ and\ \bibinfo {author} {\bibfnamefont {M.}~\bibnamefont {Dine}},\
  }\href {\doibase 10.1016/0550-3213(85)90021-5} {\bibfield  {journal}
  {\bibinfo  {journal} {Nucl.Phys.}\ }\textbf {\bibinfo {volume} {B249}},\
  \bibinfo {pages} {361} (\bibinfo {year} {1985})}\BibitemShut {NoStop}%
\bibitem [{\citenamefont {Kolb}\ and\ \citenamefont
  {Wolfram}(1980)}]{Kolb:1979qa}%
  \BibitemOpen
  \bibfield  {author} {\bibinfo {author} {\bibfnamefont {E.~W.}\ \bibnamefont
  {Kolb}}\ and\ \bibinfo {author} {\bibfnamefont {S.}~\bibnamefont {Wolfram}},\
  }\href {\doibase 10.1016/0550-3213(80)90167-4} {\bibfield  {journal}
  {\bibinfo  {journal} {Nucl.Phys.}\ }\textbf {\bibinfo {volume} {B172}},\
  \bibinfo {pages} {224} (\bibinfo {year} {1980})}\BibitemShut {NoStop}%
\bibitem [{\citenamefont {Riotto}(1998)}]{Riotto:1998bt}%
  \BibitemOpen
  \bibfield  {author} {\bibinfo {author} {\bibfnamefont {A.}~\bibnamefont
  {Riotto}},\ }\href@noop {} {\ ,\ \bibinfo {pages} {326} (\bibinfo {year}
  {1998})},\ \Eprint {http://arxiv.org/abs/hep-ph/9807454}
  {arXiv:hep-ph/9807454 [hep-ph]} \BibitemShut {NoStop}%
\bibitem [{\citenamefont {Sakharov}(1967)}]{Sakharov:1967dj}%
  \BibitemOpen
  \bibfield  {author} {\bibinfo {author} {\bibfnamefont {A.}~\bibnamefont
  {Sakharov}},\ }\href {\doibase 10.1070/PU1991v034n05ABEH002497} {\bibfield
  {journal} {\bibinfo  {journal} {Pisma Zh.Eksp.Teor.Fiz.}\ }\textbf {\bibinfo
  {volume} {5}},\ \bibinfo {pages} {32} (\bibinfo {year} {1967})}\BibitemShut
  {NoStop}%
\bibitem [{\citenamefont {Huet}\ and\ \citenamefont
  {Sather}(1995)}]{Huet:1994jb}%
  \BibitemOpen
  \bibfield  {author} {\bibinfo {author} {\bibfnamefont {P.}~\bibnamefont
  {Huet}}\ and\ \bibinfo {author} {\bibfnamefont {E.}~\bibnamefont {Sather}},\
  }\href {\doibase 10.1103/PhysRevD.51.379} {\bibfield  {journal} {\bibinfo
  {journal} {Phys.Rev.}\ }\textbf {\bibinfo {volume} {D51}},\ \bibinfo {pages}
  {379} (\bibinfo {year} {1995})},\ \Eprint
  {http://arxiv.org/abs/hep-ph/9404302} {arXiv:hep-ph/9404302 [hep-ph]}
  \BibitemShut {NoStop}%
\bibitem [{\citenamefont {Georgi}\ and\ \citenamefont
  {Glashow}(1974)}]{Georgi:1974sy}%
  \BibitemOpen
  \bibfield  {author} {\bibinfo {author} {\bibfnamefont {H.}~\bibnamefont
  {Georgi}}\ and\ \bibinfo {author} {\bibfnamefont {S.}~\bibnamefont
  {Glashow}},\ }\href {\doibase 10.1103/PhysRevLett.32.438} {\bibfield
  {journal} {\bibinfo  {journal} {Phys.Rev.Lett.}\ }\textbf {\bibinfo {volume}
  {32}},\ \bibinfo {pages} {438} (\bibinfo {year} {1974})}\BibitemShut
  {NoStop}%
\bibitem [{\citenamefont {Dimopoulos}\ and\ \citenamefont
  {Georgi}(1981)}]{Dimopoulos:1981zb}%
  \BibitemOpen
  \bibfield  {author} {\bibinfo {author} {\bibfnamefont {S.}~\bibnamefont
  {Dimopoulos}}\ and\ \bibinfo {author} {\bibfnamefont {H.}~\bibnamefont
  {Georgi}},\ }\href {\doibase 10.1016/0550-3213(81)90522-8} {\bibfield
  {journal} {\bibinfo  {journal} {Nucl.Phys.}\ }\textbf {\bibinfo {volume}
  {B193}},\ \bibinfo {pages} {150} (\bibinfo {year} {1981})}\BibitemShut
  {NoStop}%
\bibitem [{\citenamefont {Pati}\ and\ \citenamefont
  {Salam}(1974)}]{Pati:1974yy}%
  \BibitemOpen
  \bibfield  {author} {\bibinfo {author} {\bibfnamefont {J.~C.}\ \bibnamefont
  {Pati}}\ and\ \bibinfo {author} {\bibfnamefont {A.}~\bibnamefont {Salam}},\
  }\href {\doibase 10.1103/PhysRevD.10.275, 10.1103/PhysRevD.11.703.2}
  {\bibfield  {journal} {\bibinfo  {journal} {Phys.Rev.}\ }\textbf {\bibinfo
  {volume} {D10}},\ \bibinfo {pages} {275} (\bibinfo {year}
  {1974})}\BibitemShut {NoStop}%
\bibitem [{\citenamefont {Nishino}\ \emph {et~al.}(2009)\citenamefont {Nishino}
  \emph {et~al.}}]{Nishino:2009aa}%
  \BibitemOpen
  \bibfield  {author} {\bibinfo {author} {\bibfnamefont {H.}~\bibnamefont
  {Nishino}} \emph {et~al.} (\bibinfo {collaboration} {Super-Kamiokande
  Collaboration}),\ }\href {\doibase 10.1103/PhysRevLett.102.141801} {\bibfield
   {journal} {\bibinfo  {journal} {Phys.Rev.Lett.}\ }\textbf {\bibinfo {volume}
  {102}},\ \bibinfo {pages} {141801} (\bibinfo {year} {2009})},\ \Eprint
  {http://arxiv.org/abs/0903.0676} {arXiv:0903.0676 [hep-ex]} \BibitemShut
  {NoStop}%
\bibitem [{\citenamefont {Arnold}\ \emph {et~al.}(2013)\citenamefont {Arnold},
  \citenamefont {Fornal},\ and\ \citenamefont {Wise}}]{Arnold:2012sd}%
  \BibitemOpen
  \bibfield  {author} {\bibinfo {author} {\bibfnamefont {J.~M.}\ \bibnamefont
  {Arnold}}, \bibinfo {author} {\bibfnamefont {B.}~\bibnamefont {Fornal}}, \
  and\ \bibinfo {author} {\bibfnamefont {M.~B.}\ \bibnamefont {Wise}},\ }\href
  {\doibase 10.1103/PhysRevD.87.075004} {\bibfield  {journal} {\bibinfo
  {journal} {Phys.Rev.}\ }\textbf {\bibinfo {volume} {D87}},\ \bibinfo {pages}
  {075004} (\bibinfo {year} {2013})},\ \Eprint {http://arxiv.org/abs/1212.4556}
  {arXiv:1212.4556 [hep-ph]} \BibitemShut {NoStop}%
\bibitem [{\citenamefont {Babu}\ and\ \citenamefont
  {Mohapatra}(2012{\natexlab{a}})}]{Babu:2012vc}%
  \BibitemOpen
  \bibfield  {author} {\bibinfo {author} {\bibfnamefont {K.}~\bibnamefont
  {Babu}}\ and\ \bibinfo {author} {\bibfnamefont {R.}~\bibnamefont
  {Mohapatra}},\ }\href {\doibase 10.1016/j.physletb.2012.08.006} {\bibfield
  {journal} {\bibinfo  {journal} {Phys.Lett.}\ }\textbf {\bibinfo {volume}
  {B715}},\ \bibinfo {pages} {328} (\bibinfo {year} {2012}{\natexlab{a}})},\
  \Eprint {http://arxiv.org/abs/1206.5701} {arXiv:1206.5701 [hep-ph]}
  \BibitemShut {NoStop}%
\bibitem [{\citenamefont {Peggs}\ \emph {et~al.}(2009)\citenamefont {Peggs},
  \citenamefont {Calaga}, \citenamefont {Duperrier}, \citenamefont {Stovall},
  \citenamefont {Eshraqi} \emph {et~al.}}]{Peggs:2009zza}%
  \BibitemOpen
  \bibfield  {author} {\bibinfo {author} {\bibfnamefont {S.}~\bibnamefont
  {Peggs}}, \bibinfo {author} {\bibfnamefont {R.}~\bibnamefont {Calaga}},
  \bibinfo {author} {\bibfnamefont {R.}~\bibnamefont {Duperrier}}, \bibinfo
  {author} {\bibfnamefont {J.}~\bibnamefont {Stovall}}, \bibinfo {author}
  {\bibfnamefont {M.}~\bibnamefont {Eshraqi}},  \emph {et~al.},\ }\href@noop {}
  {\  (\bibinfo {year} {2009})}\BibitemShut {NoStop}%
\bibitem [{\citenamefont {Richardson}\ and\ \citenamefont
  {Winn}(2012)}]{Richardson:2011df}%
  \BibitemOpen
  \bibfield  {author} {\bibinfo {author} {\bibfnamefont {P.}~\bibnamefont
  {Richardson}}\ and\ \bibinfo {author} {\bibfnamefont {D.}~\bibnamefont
  {Winn}},\ }\href {\doibase 10.1140/epjc/s10052-012-1862-z} {\bibfield
  {journal} {\bibinfo  {journal} {Eur.Phys.J.}\ }\textbf {\bibinfo {volume}
  {C72}},\ \bibinfo {pages} {1862} (\bibinfo {year} {2012})},\ \Eprint
  {http://arxiv.org/abs/1108.6154} {arXiv:1108.6154 [hep-ph]} \BibitemShut
  {NoStop}%
\bibitem [{\citenamefont {Han}\ \emph {et~al.}(2010)\citenamefont {Han},
  \citenamefont {Lewis},\ and\ \citenamefont {McElmurry}}]{Han:2009ya}%
  \BibitemOpen
  \bibfield  {author} {\bibinfo {author} {\bibfnamefont {T.}~\bibnamefont
  {Han}}, \bibinfo {author} {\bibfnamefont {I.}~\bibnamefont {Lewis}}, \ and\
  \bibinfo {author} {\bibfnamefont {T.}~\bibnamefont {McElmurry}},\ }\href
  {\doibase 10.1007/JHEP01(2010)123} {\bibfield  {journal} {\bibinfo  {journal}
  {JHEP}\ }\textbf {\bibinfo {volume} {1001}},\ \bibinfo {pages} {123}
  (\bibinfo {year} {2010})},\ \Eprint {http://arxiv.org/abs/0909.2666}
  {arXiv:0909.2666 [hep-ph]} \BibitemShut {NoStop}%
\bibitem [{\citenamefont {Kilian}\ \emph {et~al.}(2012)\citenamefont {Kilian},
  \citenamefont {Ohl}, \citenamefont {Reuter},\ and\ \citenamefont
  {Speckner}}]{Kilian:2012pz}%
  \BibitemOpen
  \bibfield  {author} {\bibinfo {author} {\bibfnamefont {W.}~\bibnamefont
  {Kilian}}, \bibinfo {author} {\bibfnamefont {T.}~\bibnamefont {Ohl}},
  \bibinfo {author} {\bibfnamefont {J.}~\bibnamefont {Reuter}}, \ and\ \bibinfo
  {author} {\bibfnamefont {C.}~\bibnamefont {Speckner}},\ }\href {\doibase
  10.1007/JHEP10(2012)022} {\bibfield  {journal} {\bibinfo  {journal} {JHEP}\
  }\textbf {\bibinfo {volume} {1210}},\ \bibinfo {pages} {022} (\bibinfo {year}
  {2012})},\ \Eprint {http://arxiv.org/abs/1206.3700} {arXiv:1206.3700
  [hep-ph]} \BibitemShut {NoStop}%
\bibitem [{\citenamefont {Kolb}\ and\ \citenamefont
  {Turner}(1990)}]{Kolb:1990vq}%
  \BibitemOpen
  \bibfield  {author} {\bibinfo {author} {\bibfnamefont {E.~W.}\ \bibnamefont
  {Kolb}}\ and\ \bibinfo {author} {\bibfnamefont {M.~S.}\ \bibnamefont
  {Turner}},\ }\href@noop {} {\bibfield  {journal} {\bibinfo  {journal}
  {Front.Phys.}\ }\textbf {\bibinfo {volume} {69}},\ \bibinfo {pages} {1}
  (\bibinfo {year} {1990})}\BibitemShut {NoStop}%
\bibitem [{\citenamefont {'t~Hooft}(1976)}]{'tHooft:1976fv}%
  \BibitemOpen
  \bibfield  {author} {\bibinfo {author} {\bibfnamefont {G.}~\bibnamefont
  {'t~Hooft}},\ }\href {\doibase 10.1103/PhysRevD.18.2199.3,
  10.1103/PhysRevD.14.3432} {\bibfield  {journal} {\bibinfo  {journal}
  {Phys.Rev.}\ }\textbf {\bibinfo {volume} {D14}},\ \bibinfo {pages} {3432}
  (\bibinfo {year} {1976})}\BibitemShut {NoStop}%
\bibitem [{\citenamefont {Arnold}\ and\ \citenamefont
  {McLerran}(1987)}]{Arnold:1987mh}%
  \BibitemOpen
  \bibfield  {author} {\bibinfo {author} {\bibfnamefont {P.~B.}\ \bibnamefont
  {Arnold}}\ and\ \bibinfo {author} {\bibfnamefont {L.~D.}\ \bibnamefont
  {McLerran}},\ }\href {\doibase 10.1103/PhysRevD.36.581} {\bibfield  {journal}
  {\bibinfo  {journal} {Phys.Rev.}\ }\textbf {\bibinfo {volume} {D36}},\
  \bibinfo {pages} {581} (\bibinfo {year} {1987})}\BibitemShut {NoStop}%
\bibitem [{\citenamefont {Greene}\ \emph {et~al.}(1997)\citenamefont {Greene},
  \citenamefont {Kofman}, \citenamefont {Linde},\ and\ \citenamefont
  {Starobinsky}}]{Greene:1997fu}%
  \BibitemOpen
  \bibfield  {author} {\bibinfo {author} {\bibfnamefont {P.~B.}\ \bibnamefont
  {Greene}}, \bibinfo {author} {\bibfnamefont {L.}~\bibnamefont {Kofman}},
  \bibinfo {author} {\bibfnamefont {A.~D.}\ \bibnamefont {Linde}}, \ and\
  \bibinfo {author} {\bibfnamefont {A.~A.}\ \bibnamefont {Starobinsky}},\
  }\href {\doibase 10.1103/PhysRevD.56.6175} {\bibfield  {journal} {\bibinfo
  {journal} {Phys.Rev.}\ }\textbf {\bibinfo {volume} {D56}},\ \bibinfo {pages}
  {6175} (\bibinfo {year} {1997})},\ \Eprint
  {http://arxiv.org/abs/hep-ph/9705347} {arXiv:hep-ph/9705347 [hep-ph]}
  \BibitemShut {NoStop}%
\bibitem [{\citenamefont {Giudice}\ \emph {et~al.}(2004)\citenamefont
  {Giudice}, \citenamefont {Notari}, \citenamefont {Raidal}, \citenamefont
  {Riotto},\ and\ \citenamefont {Strumia}}]{Giudice:2003jh}%
  \BibitemOpen
  \bibfield  {author} {\bibinfo {author} {\bibfnamefont {G.}~\bibnamefont
  {Giudice}}, \bibinfo {author} {\bibfnamefont {A.}~\bibnamefont {Notari}},
  \bibinfo {author} {\bibfnamefont {M.}~\bibnamefont {Raidal}}, \bibinfo
  {author} {\bibfnamefont {A.}~\bibnamefont {Riotto}}, \ and\ \bibinfo {author}
  {\bibfnamefont {A.}~\bibnamefont {Strumia}},\ }\href {\doibase
  10.1016/j.nuclphysb.2004.02.019} {\bibfield  {journal} {\bibinfo  {journal}
  {Nucl.Phys.}\ }\textbf {\bibinfo {volume} {B685}},\ \bibinfo {pages} {89}
  (\bibinfo {year} {2004})},\ \Eprint {http://arxiv.org/abs/hep-ph/0310123}
  {arXiv:hep-ph/0310123 [hep-ph]} \BibitemShut {NoStop}%
\bibitem [{\citenamefont {Strumia}(2006)}]{Strumia:2006qk}%
  \BibitemOpen
  \bibfield  {author} {\bibinfo {author} {\bibfnamefont {A.}~\bibnamefont
  {Strumia}},\ }\href@noop {} {\ ,\ \bibinfo {pages} {655} (\bibinfo {year}
  {2006})},\ \Eprint {http://arxiv.org/abs/hep-ph/0608347}
  {arXiv:hep-ph/0608347 [hep-ph]} \BibitemShut {NoStop}%
\bibitem [{\citenamefont {Buchmuller}\ \emph {et~al.}(2005)\citenamefont
  {Buchmuller}, \citenamefont {Di~Bari},\ and\ \citenamefont
  {Plumacher}}]{Buchmuller:2004nz}%
  \BibitemOpen
  \bibfield  {author} {\bibinfo {author} {\bibfnamefont {W.}~\bibnamefont
  {Buchmuller}}, \bibinfo {author} {\bibfnamefont {P.}~\bibnamefont {Di~Bari}},
  \ and\ \bibinfo {author} {\bibfnamefont {M.}~\bibnamefont {Plumacher}},\
  }\href {\doibase 10.1016/j.aop.2004.02.003} {\bibfield  {journal} {\bibinfo
  {journal} {Annals Phys.}\ }\textbf {\bibinfo {volume} {315}},\ \bibinfo
  {pages} {305} (\bibinfo {year} {2005})},\ \Eprint
  {http://arxiv.org/abs/hep-ph/0401240} {arXiv:hep-ph/0401240 [hep-ph]}
  \BibitemShut {NoStop}%
\bibitem [{\citenamefont {Chen}\ \emph {et~al.}(2009)\citenamefont {Chen},
  \citenamefont {Klemm}, \citenamefont {Rentala},\ and\ \citenamefont
  {Wang}}]{Chen:2008hh}%
  \BibitemOpen
  \bibfield  {author} {\bibinfo {author} {\bibfnamefont {C.-R.}\ \bibnamefont
  {Chen}}, \bibinfo {author} {\bibfnamefont {W.}~\bibnamefont {Klemm}},
  \bibinfo {author} {\bibfnamefont {V.}~\bibnamefont {Rentala}}, \ and\
  \bibinfo {author} {\bibfnamefont {K.}~\bibnamefont {Wang}},\ }\href {\doibase
  10.1103/PhysRevD.79.054002} {\bibfield  {journal} {\bibinfo  {journal}
  {Phys.Rev.}\ }\textbf {\bibinfo {volume} {D79}},\ \bibinfo {pages} {054002}
  (\bibinfo {year} {2009})},\ \Eprint {http://arxiv.org/abs/0811.2105}
  {arXiv:0811.2105 [hep-ph]} \BibitemShut {NoStop}%
\bibitem [{\citenamefont {Manohar}\ and\ \citenamefont
  {Wise}(2006)}]{Manohar:2006ga}%
  \BibitemOpen
  \bibfield  {author} {\bibinfo {author} {\bibfnamefont {A.~V.}\ \bibnamefont
  {Manohar}}\ and\ \bibinfo {author} {\bibfnamefont {M.~B.}\ \bibnamefont
  {Wise}},\ }\href {\doibase 10.1103/PhysRevD.74.035009} {\bibfield  {journal}
  {\bibinfo  {journal} {Phys.Rev.}\ }\textbf {\bibinfo {volume} {D74}},\
  \bibinfo {pages} {035009} (\bibinfo {year} {2006})},\ \Eprint
  {http://arxiv.org/abs/hep-ph/0606172} {arXiv:hep-ph/0606172 [hep-ph]}
  \BibitemShut {NoStop}%
\bibitem [{\citenamefont {Bennett}\ \emph {et~al.}(2013)\citenamefont {Bennett}
  \emph {et~al.}}]{Bennett:2012zja}%
  \BibitemOpen
  \bibfield  {author} {\bibinfo {author} {\bibfnamefont {C.}~\bibnamefont
  {Bennett}} \emph {et~al.} (\bibinfo {collaboration} {WMAP}),\ }\href
  {\doibase 10.1088/0067-0049/208/2/20} {\bibfield  {journal} {\bibinfo
  {journal} {Astrophys.J.Suppl.}\ }\textbf {\bibinfo {volume} {208}},\ \bibinfo
  {pages} {20} (\bibinfo {year} {2013})},\ \Eprint
  {http://arxiv.org/abs/1212.5225} {arXiv:1212.5225 [astro-ph.CO]} \BibitemShut
  {NoStop}%
\bibitem [{\citenamefont {Babu}\ and\ \citenamefont
  {Mohapatra}(2012{\natexlab{b}})}]{Babu:2012iv}%
  \BibitemOpen
  \bibfield  {author} {\bibinfo {author} {\bibfnamefont {K.}~\bibnamefont
  {Babu}}\ and\ \bibinfo {author} {\bibfnamefont {R.}~\bibnamefont
  {Mohapatra}},\ }\href {\doibase 10.1103/PhysRevLett.109.091803} {\bibfield
  {journal} {\bibinfo  {journal} {Phys.Rev.Lett.}\ }\textbf {\bibinfo {volume}
  {109}},\ \bibinfo {pages} {091803} (\bibinfo {year} {2012}{\natexlab{b}})},\
  \Eprint {http://arxiv.org/abs/1207.5771} {arXiv:1207.5771 [hep-ph]}
  \BibitemShut {NoStop}%
\bibitem [{\citenamefont {Kamyshkov}(2002)}]{Kamyshkov:2002vm}%
  \BibitemOpen
  \bibfield  {author} {\bibinfo {author} {\bibfnamefont {Y.~A.}\ \bibnamefont
  {Kamyshkov}},\ }\href@noop {} {\  (\bibinfo {year} {2002})},\ \Eprint
  {http://arxiv.org/abs/hep-ex/0211006} {arXiv:hep-ex/0211006 [hep-ex]}
  \BibitemShut {NoStop}%
\bibitem [{\citenamefont {Mohapatra}(2009)}]{Mohapatra:2009wp}%
  \BibitemOpen
  \bibfield  {author} {\bibinfo {author} {\bibfnamefont {R.}~\bibnamefont
  {Mohapatra}},\ }\href {\doibase 10.1088/0954-3899/36/10/104006} {\bibfield
  {journal} {\bibinfo  {journal} {J.Phys.}\ }\textbf {\bibinfo {volume}
  {G36}},\ \bibinfo {pages} {104006} (\bibinfo {year} {2009})},\ \Eprint
  {http://arxiv.org/abs/0902.0834} {arXiv:0902.0834 [hep-ph]} \BibitemShut
  {NoStop}%
\bibitem [{\citenamefont {Kuzmin}(1970)}]{Kuzmin:1970nx}%
  \BibitemOpen
  \bibfield  {author} {\bibinfo {author} {\bibfnamefont {V.}~\bibnamefont
  {Kuzmin}},\ }\href@noop {} {\bibfield  {journal} {\bibinfo  {journal} {Pisma
  Zh.Eksp.Teor.Fiz.}\ }\textbf {\bibinfo {volume} {12}},\ \bibinfo {pages}
  {335} (\bibinfo {year} {1970})}\BibitemShut {NoStop}%
\bibitem [{\citenamefont {Chacko}\ and\ \citenamefont
  {Mohapatra}(1999)}]{Chacko:1998td}%
  \BibitemOpen
  \bibfield  {author} {\bibinfo {author} {\bibfnamefont {Z.}~\bibnamefont
  {Chacko}}\ and\ \bibinfo {author} {\bibfnamefont {R.}~\bibnamefont
  {Mohapatra}},\ }\href {\doibase 10.1103/PhysRevD.59.055004} {\bibfield
  {journal} {\bibinfo  {journal} {Phys.Rev.}\ }\textbf {\bibinfo {volume}
  {D59}},\ \bibinfo {pages} {055004} (\bibinfo {year} {1999})},\ \Eprint
  {http://arxiv.org/abs/hep-ph/9802388} {arXiv:hep-ph/9802388 [hep-ph]}
  \BibitemShut {NoStop}%
\bibitem [{\citenamefont {Dutta}\ \emph {et~al.}(2006)\citenamefont {Dutta},
  \citenamefont {Mimura},\ and\ \citenamefont {Mohapatra}}]{Dutta:2005af}%
  \BibitemOpen
  \bibfield  {author} {\bibinfo {author} {\bibfnamefont {B.}~\bibnamefont
  {Dutta}}, \bibinfo {author} {\bibfnamefont {Y.}~\bibnamefont {Mimura}}, \
  and\ \bibinfo {author} {\bibfnamefont {R.}~\bibnamefont {Mohapatra}},\ }\href
  {\doibase 10.1103/PhysRevLett.96.061801} {\bibfield  {journal} {\bibinfo
  {journal} {Phys.Rev.Lett.}\ }\textbf {\bibinfo {volume} {96}},\ \bibinfo
  {pages} {061801} (\bibinfo {year} {2006})},\ \Eprint
  {http://arxiv.org/abs/hep-ph/0510291} {arXiv:hep-ph/0510291 [hep-ph]}
  \BibitemShut {NoStop}%
\bibitem [{\citenamefont {Tsutsui}\ \emph {et~al.}(2004)\citenamefont {Tsutsui}
  \emph {et~al.}}]{Tsutsui:2004qc}%
  \BibitemOpen
  \bibfield  {author} {\bibinfo {author} {\bibfnamefont {N.}~\bibnamefont
  {Tsutsui}} \emph {et~al.} (\bibinfo {collaboration} {CP-PACS Collaboration,
  JLQCD Collaborations}),\ }\href {\doibase 10.1103/PhysRevD.70.111501}
  {\bibfield  {journal} {\bibinfo  {journal} {Phys.Rev.}\ }\textbf {\bibinfo
  {volume} {D70}},\ \bibinfo {pages} {111501} (\bibinfo {year} {2004})},\
  \Eprint {http://arxiv.org/abs/hep-lat/0402026} {arXiv:hep-lat/0402026
  [hep-lat]} \BibitemShut {NoStop}%
\bibitem [{\citenamefont {Abe}\ \emph {et~al.}(2011)\citenamefont {Abe} \emph
  {et~al.}}]{Abe:2011ky}%
  \BibitemOpen
  \bibfield  {author} {\bibinfo {author} {\bibfnamefont {K.}~\bibnamefont
  {Abe}} \emph {et~al.} (\bibinfo {collaboration} {Super-Kamiokande
  Collaboration}),\ }\href@noop {} {\  (\bibinfo {year} {2011})},\ \Eprint
  {http://arxiv.org/abs/1109.4227} {arXiv:1109.4227 [hep-ex]} \BibitemShut
  {NoStop}%
\bibitem [{\citenamefont {Babu}\ \emph {et~al.}(2013)\citenamefont {Babu},
  \citenamefont {Banerjee}, \citenamefont {Baxter}, \citenamefont {Berezhiani},
  \citenamefont {Bergevin} \emph {et~al.}}]{Babu:2013yww}%
  \BibitemOpen
  \bibfield  {author} {\bibinfo {author} {\bibfnamefont {K.}~\bibnamefont
  {Babu}}, \bibinfo {author} {\bibfnamefont {S.}~\bibnamefont {Banerjee}},
  \bibinfo {author} {\bibfnamefont {D.}~\bibnamefont {Baxter}}, \bibinfo
  {author} {\bibfnamefont {Z.}~\bibnamefont {Berezhiani}}, \bibinfo {author}
  {\bibfnamefont {M.}~\bibnamefont {Bergevin}},  \emph {et~al.},\ }\href@noop
  {} {\  (\bibinfo {year} {2013})},\ \Eprint {http://arxiv.org/abs/1310.8593}
  {arXiv:1310.8593 [hep-ex]} \BibitemShut {NoStop}%
\bibitem [{\citenamefont {Czarnecki}\ and\ \citenamefont
  {Krause}(1997)}]{Czarnecki:1997bu}%
  \BibitemOpen
  \bibfield  {author} {\bibinfo {author} {\bibfnamefont {A.}~\bibnamefont
  {Czarnecki}}\ and\ \bibinfo {author} {\bibfnamefont {B.}~\bibnamefont
  {Krause}},\ }\href {\doibase 10.1103/PhysRevLett.78.4339} {\bibfield
  {journal} {\bibinfo  {journal} {Phys.Rev.Lett.}\ }\textbf {\bibinfo {volume}
  {78}},\ \bibinfo {pages} {4339} (\bibinfo {year} {1997})},\ \Eprint
  {http://arxiv.org/abs/hep-ph/9704355} {arXiv:hep-ph/9704355 [hep-ph]}
  \BibitemShut {NoStop}%
\bibitem [{\citenamefont {Mannel}\ and\ \citenamefont
  {Uraltsev}(2012)}]{Mannel:2012qk}%
  \BibitemOpen
  \bibfield  {author} {\bibinfo {author} {\bibfnamefont {T.}~\bibnamefont
  {Mannel}}\ and\ \bibinfo {author} {\bibfnamefont {N.}~\bibnamefont
  {Uraltsev}},\ }\href {\doibase 10.1103/PhysRevD.85.096002} {\bibfield
  {journal} {\bibinfo  {journal} {Phys.Rev.}\ }\textbf {\bibinfo {volume}
  {D85}},\ \bibinfo {pages} {096002} (\bibinfo {year} {2012})},\ \Eprint
  {http://arxiv.org/abs/1202.6270} {arXiv:1202.6270 [hep-ph]} \BibitemShut
  {NoStop}%
\bibitem [{\citenamefont {Serebrov}\ \emph {et~al.}(2013)\citenamefont
  {Serebrov}, \citenamefont {Kolomenskiy}, \citenamefont {Pirozhkov},
  \citenamefont {Krasnoshekova}, \citenamefont {Vasiliev} \emph
  {et~al.}}]{Serebrov:2013tba}%
  \BibitemOpen
  \bibfield  {author} {\bibinfo {author} {\bibfnamefont {A.}~\bibnamefont
  {Serebrov}}, \bibinfo {author} {\bibfnamefont {E.}~\bibnamefont
  {Kolomenskiy}}, \bibinfo {author} {\bibfnamefont {A.}~\bibnamefont
  {Pirozhkov}}, \bibinfo {author} {\bibfnamefont {I.}~\bibnamefont
  {Krasnoshekova}}, \bibinfo {author} {\bibfnamefont {A.}~\bibnamefont
  {Vasiliev}},  \emph {et~al.},\ }\href@noop {} {\  (\bibinfo {year} {2013})},\
  \Eprint {http://arxiv.org/abs/1310.5588} {arXiv:1310.5588 [nucl-ex]}
  \BibitemShut {NoStop}%
\bibitem [{\citenamefont {Mohapatra}\ \emph {et~al.}(2008)\citenamefont
  {Mohapatra}, \citenamefont {Okada},\ and\ \citenamefont
  {Yu}}]{Mohapatra:2007af}%
  \BibitemOpen
  \bibfield  {author} {\bibinfo {author} {\bibfnamefont {R.}~\bibnamefont
  {Mohapatra}}, \bibinfo {author} {\bibfnamefont {N.}~\bibnamefont {Okada}}, \
  and\ \bibinfo {author} {\bibfnamefont {H.-B.}\ \bibnamefont {Yu}},\ }\href
  {\doibase 10.1103/PhysRevD.77.011701} {\bibfield  {journal} {\bibinfo
  {journal} {Phys.Rev.}\ }\textbf {\bibinfo {volume} {D77}},\ \bibinfo {pages}
  {011701} (\bibinfo {year} {2008})},\ \Eprint {http://arxiv.org/abs/0709.1486}
  {arXiv:0709.1486 [hep-ph]} \BibitemShut {NoStop}%
\bibitem [{\citenamefont {Berger}\ \emph {et~al.}(2010)\citenamefont {Berger},
  \citenamefont {Cao}, \citenamefont {Chen}, \citenamefont {Shaughnessy},\ and\
  \citenamefont {Zhang}}]{Berger:2010fy}%
  \BibitemOpen
  \bibfield  {author} {\bibinfo {author} {\bibfnamefont {E.~L.}\ \bibnamefont
  {Berger}}, \bibinfo {author} {\bibfnamefont {Q.-H.}\ \bibnamefont {Cao}},
  \bibinfo {author} {\bibfnamefont {C.-R.}\ \bibnamefont {Chen}}, \bibinfo
  {author} {\bibfnamefont {G.}~\bibnamefont {Shaughnessy}}, \ and\ \bibinfo
  {author} {\bibfnamefont {H.}~\bibnamefont {Zhang}},\ }\href {\doibase
  10.1103/PhysRevLett.105.181802} {\bibfield  {journal} {\bibinfo  {journal}
  {Phys.Rev.Lett.}\ }\textbf {\bibinfo {volume} {105}},\ \bibinfo {pages}
  {181802} (\bibinfo {year} {2010})},\ \Eprint {http://arxiv.org/abs/1005.2622}
  {arXiv:1005.2622 [hep-ph]} \BibitemShut {NoStop}%
\bibitem [{\citenamefont {Zhan}\ \emph {et~al.}(2014)\citenamefont {Zhan},
  \citenamefont {Liu}, \citenamefont {Li}, \citenamefont {Li},\ and\
  \citenamefont {Li}}]{Zhan:2013sza}%
  \BibitemOpen
  \bibfield  {author} {\bibinfo {author} {\bibfnamefont {Y.~C.}\ \bibnamefont
  {Zhan}}, \bibinfo {author} {\bibfnamefont {Z.~L.}\ \bibnamefont {Liu}},
  \bibinfo {author} {\bibfnamefont {S.~A.}\ \bibnamefont {Li}}, \bibinfo
  {author} {\bibfnamefont {C.~S.}\ \bibnamefont {Li}}, \ and\ \bibinfo {author}
  {\bibfnamefont {H.~T.}\ \bibnamefont {Li}},\ }\href {\doibase
  10.1140/epjc/s10052-014-2716-7} {\bibfield  {journal} {\bibinfo  {journal}
  {Eur.Phys.J.}\ }\textbf {\bibinfo {volume} {C74}},\ \bibinfo {pages} {2716}
  (\bibinfo {year} {2014})},\ \Eprint {http://arxiv.org/abs/1305.5152}
  {arXiv:1305.5152 [hep-ph]} \BibitemShut {NoStop}%
\end{thebibliography}%

\end{document}